\documentclass[11pt]{article}

\usepackage[preprint]{acl}

\usepackage{times}
\usepackage{latexsym}
\usepackage{todonotes}

\usepackage[T1]{fontenc}

\usepackage[utf8]{inputenc}

\usepackage{microtype}

\usepackage{inconsolata}

\usepackage{graphicx}
\usepackage{natbib}
\usepackage{amsmath}
\usepackage{amsfonts} 
\usepackage[linesnumbered,ruled,vlined]{algorithm2e}
\usepackage{multirow}
\usepackage{makecell}
\usepackage{soul}
\usepackage{booktabs}
\usepackage{enumitem}

\SetKwInOut{KwIn}{Input}  
\SetKwInOut{KwOut}{Output} 

\title{DenoiseRank: Learning to Rank by Diffusion Models}

\author{
  \hspace{1em} Ying Wang \hspace{3em} Preslav Nakov \hspace{1em} Shangsong Liang$^\ast$ \\
  Sun Yat-sen University \hspace{1em} MBZUAI \hspace{1em} Sun Yat-sen University \\
  Guangzhou, China \hspace{1em} Abu Dhabi, UAE \hspace{1em} Guangzhou, China \\
  \texttt{\{wangy2593, liangshangsong\}@gmail.com} \hspace{0.5em} \texttt{preslav.nakov@mbzuai.ac.ae}
}

\begin{document}
\maketitle
\begin{abstract}
Learning to rank (LTR) is one of the core tasks in Machine Learning. Traditional LTR models have made great progress, but nearly all of them are implemented from discriminative perspective. In this paper, we aim at addressing LTR from a novel perspective, i.e., by a deep generative model. Specifically, we propose a novel denoise rank model, DenoiseRank, which noises the relevant labels in the diffusion process and denoises them  on the query documents in the reverse process to accurately predict their distribution. Our model is the first to address traditional LTR from generative perspective and is a diffusion method for LTR. Our extensive experiments on benchmark datasets demonstrated the effectiveness of DenoiseRank, and we believe it provides a benchmark for generative LTR task. Source code of our model is publicly available from: \url{http://anonymous.for.review}
\end{abstract}


\section{Introduction}
\label{sec:Introduction}

Learning to Rank (LTR) is one of the core tasks in Information Retrieval (IR) addressed by supervised machine learning techniques that are used to automatically construct ranking models. Its primary goal is to rank items (documents, products, etc.) by relevance for a given query. LTR has been commonly used in a wide spectrum of IR applications, e.g., recommender systems~\citep{wu2021unbiased} and question-answering~\citep{yang2024kg,khamnuansin2024mrrank}. Traditional LTR (T-LTR) methods (refers to algorithms that mainly involve handcrafted feature ranking) can be classified into two types, tree-based models~\citep{lucchese2025explainable,chen2024ms} and neural-based LTR models~\citep{jin2024inforank,argouarc2024generative,padhye2023deep,jagerman2022optimizing,pang2020setrank}. Such approaches mainly follow a discriminative paradigm to rank \citep{chen2024ms,argouarc2024generative,gu2020deep,ke2017lightgbm,qin2021neural,pang2020setrank}, which struggle with complex distributions of user feedback that include a lot of noise, limiting their ability for uncertainty estimation and produce robust rankings. Beside, T-LTR tends to rank consistently overtime that unable to provide a fair ranking. 


Recently, generative models have shown strong distribution-fitting and diversity generation ability for many tasks such as machine translation~\citep{yuasa2023multimodal}, text classification~\citep{yuan2024roic} and sentiment analysis~\citep{yin2024textgt,zhang2024reconstructing}.  However, challenges arise when adapting these generative paradigms to ranking tasks. GANs face mode collapse and training instability issue, VAEs suffer from posterior collapse issue, and autoregressive LMs \cite{tamber2025lit,li2024learning,tamber2023scaling} are computationally heavy and conceptually different from T-LTR. (More discussions in Appendix \ref{sec:Motivation})

Of special interest to us is Denoising Diffusion Probabilistic Models (DDPMs)~\citep{ho2020denoising,nichol2021improved}, which have demonstrated significant potential in text generation ~\citep{yi2024diffusion} and conversation system~\citep{mughal2024convofusion,chen2023controllable} in recent years. DDPMs are particularly promising for LTR due to their ability to learn conditional probability distribution, uncertainty estimation, diversity representations, and training stability. Accordingly, we ask: \textit{Can DDPMs effectively estimate the conditional distribution $P(Y|D)$ for T-LTR (given a ranked list of documents $D$ in response to an input query), leading to better and diverse ranking?}

Specifically, we 
propose a novel denoising ranking model, DenoiseRank, to address the T-LTR task. Our DenoiseRank is a DDPM (Sec.\ref{sec:DenoiseRankingModel}) model and consists of three main components: diffusion process, reverse process and denoise neural network. In our proposed DenoiseRank, we first gradually inject Gaussian noise into the input corresponding labels $Y$ within a number of timesteps, (based on the Markov process \citep{sohl2015deep}), and consequently $Y$ becomes an isotropic Gaussian distribution. Then, noised labels are fed into the denoise neural network which will output denoised labels. The forward and reverse processes are similar to making an uncertainty estimate based on user feedback. We repeat this process on datasets and optimize the model with a special loss function, in order to learn the conditional probability distribution $P(Y|D)$. Note that the Denoise neural network mainly consists of a custom feedforward network and a Transformer-Encoder network. Finally, during the reverse process, the labels $Y$ are randomly sampled from Gaussian noise and fed into the well-trained Denoise neural network. After enough iterations, accurate labels are obtained. We demonstrate the effectiveness of the proposed DenoiseRank through extensive experimentation (Sec.~\ref{sec:Experiments}) on the popular LTR dataset \citep{qin2013introducing,chapelle2011yahoo,dato2016fast} and analyse its characteristics in ablation studies (Appendix.~\ref{sec:AblationStudy}). Experimental results show that DenoiseRank  either outperforms or performs equivalently to recent LTR models \citep{buyl2023rankformer,qin2021neural}. Our DenoiseRank can be utilised as a benchmark for future neural ranking models and can be extended to other sequence prediction tasks, such as ``Multi-Objective Ranking (MOR) learning'' \citep{gu2020deep}.

The main contributions of this paper are:
\begin{enumerate}[topsep=0pt,itemsep=-1ex,partopsep=1ex,parsep=1ex]
    \item We study the problem of T-LTR from a generative perspective and introduce a diffusion model solution for the first time.
    \item We propose a novel DenoiseRank model for T-LTR, which can be used as a benchmark for future generative neural ranking models.
    \item Extensive experiments were conducted on benchmark datasets to demonstrate the effectiveness of DenoiseRank, and optimal performance was achieved against most metrics.
    \item We introduce a new metric, $RSD@(K,M)$, to evaluate the ability of the model to produce diverse ranked lists. Our DenoiseRank has been proven to rank documents diversely.
\end{enumerate}


\section{Definition of the LTR Task}
\label{sec:background}



We denote a training dataset as query set $Q=\{q_l\}_{l=1}^L$ and their corresponding document and label set $\{(D_l, Y_l)\mid q_l\}_{l=1}^L$, where $D_l$ is a document list that contains $n$ documents $D_{l,i}$ to be sorted, $i \in [1,n]$, $D_{l,i} \in \mathbb{R}^k $, and $k$ is the dimensions of the feature. $Y_l$ is the label list for the corresponding documents list $D_l$, with $Y_{l,i}>0$ indicating the document $D_{l,i}$ being relevant to the query and $Y_{l, i}=0$ otherwise. 
$L$ denotes the total number of queries contained in the dataset. The goal of LTR is to train a ranking function $f(Q,D)$, which can be used to accurately predict the relevance score of documents. We approximate the ranking function by training a model optimized by the loss $\mathcal{L}(f; Q, D)$. 


\section{Denoise Ranking Model}
\label{sec:DenoiseRankingModel}

In this section, we first provide an overview of our proposed DenoiseRank model. 
We then detail our DenoiseRank model, including the diffusion process and the inverse process. Finally we provide the training and inference algorithm and discuss the advance of our model compare to the others. We provide further details on the Denoise Neural Network of DenoiseRank in Appendix \ref{sec:DenoiseRankModelArchitecture}.

\begin{figure*}[!t]
  \includegraphics[width=0.5\linewidth]{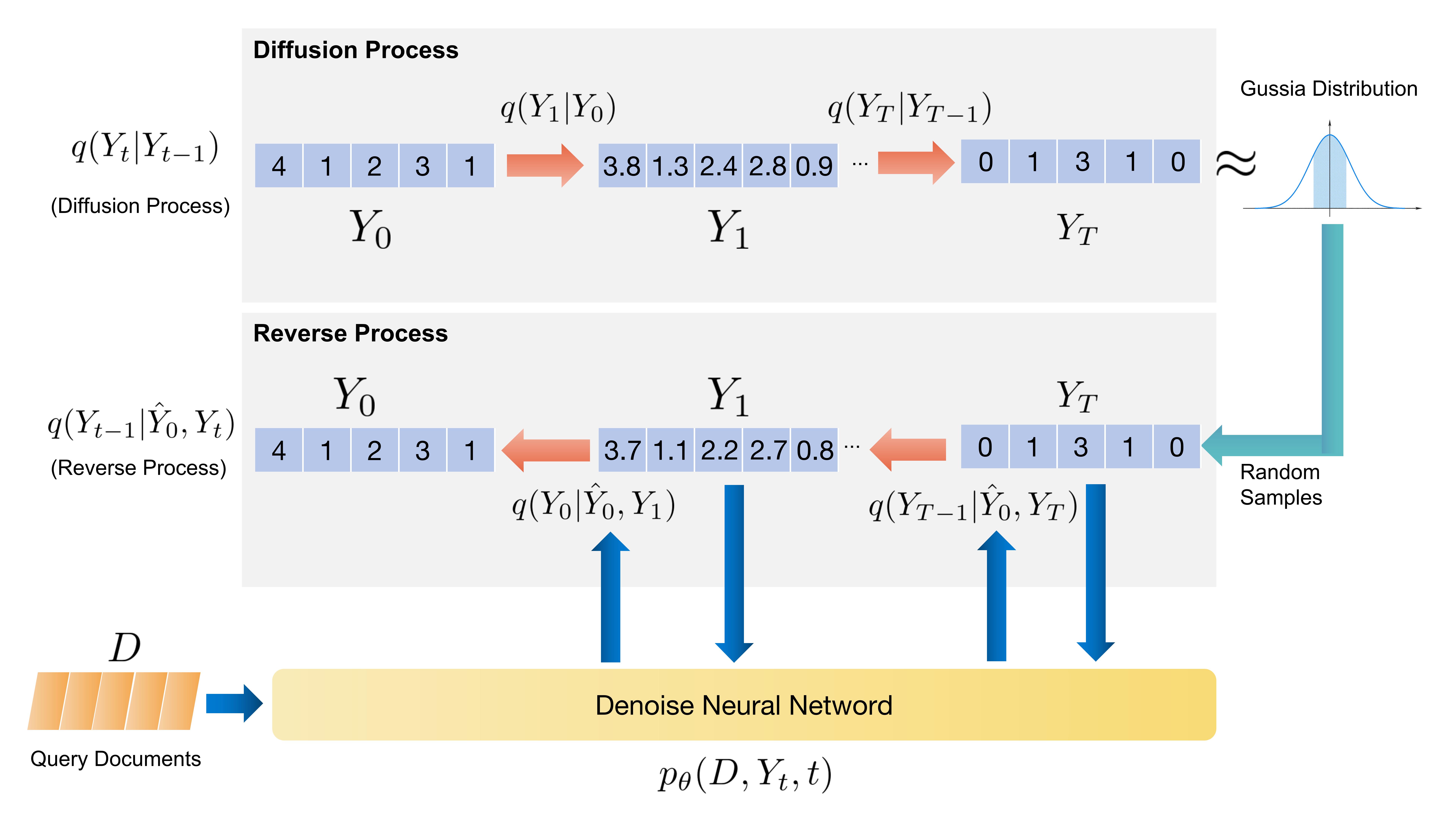} \hfill
  \includegraphics[width=0.5\linewidth]{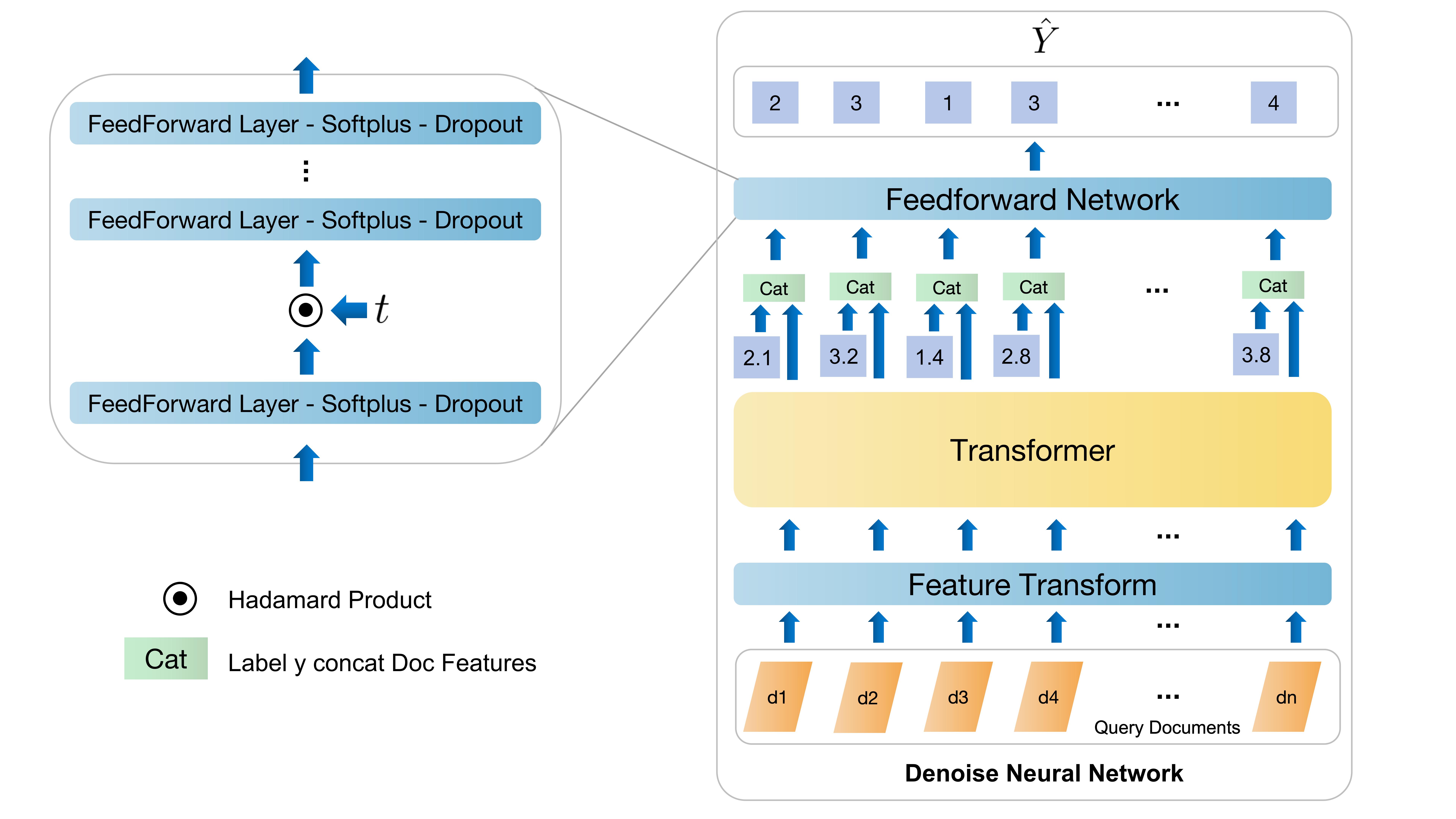}
  \caption {The left panel illustrates the diffusion and reverse processes in DenoiseRank, while the right panel shows the architecture of its denoising neural network. The FeedForward Layer is a linear layer, softplus is the activation function, and dropout is applied as a regularization layer. Feature transformation is applied when document features are input into the transformer block.}
  \label{fig:DenoiseRank_Diffusion}
\end{figure*}

\subsection{Overview of Our DenoiseRank}
The overview of our proposed DenoiseRank is provided in Figure~\ref{fig:DenoiseRank_Diffusion}. As shown in the figure, our DenoiseRank is built based on DDPM and consists of three components: Diffusion Process, reverse process and denoise neural network. It first takes documents $D$ in response to a given query and corresponding labels $Y$ as input, then the Gaussian noise are injected into the labels through diffusion process. Next, noised labels and documents are fed into the denoise neural network to train and optimize them. Finally, in the reverse process, labels from randomly sampling and the corresponding documents are fed into the well-trained denoise neural network to predict the ground-truth labels step by step. Note that the Denoise Neural Network in DenoiseRank is implemented by a custom feedforward network and a Transformer-Encoder network, see Figure~\ref{fig:DenoiseRank_Diffusion} on the right.

\subsection{The DenoiseRank Model}

As shown in Figure \ref{fig:DenoiseRank_Diffusion}, given are a list of  documents $D={d_1,d_2,...dn}$ in response to a query, and the corresponding list of feedback labels $Y$, $Y={y_1,y_2,...y_n}$. We hope that given $D$, denoiseRank will be able to predict the list of relevance labels $Y$ correctly through the reverse diffusion process. Our model is trained to approximate the distribution $p(Y|D)$. 
Our goal of training can be formulated as $p_{\theta}(Y_0|D) := \int p_{\theta}(Y_{0:T}|D)dY_{1:T}$, where $Y_0$ is the input labels $Y$ noised at timestep $0$, $Y_0,...,Y_T$ is noising data sampled from $Y_0 \sim q(Y_0)$. $p_{\theta}(Y_{0:T}|D)$ is the reverse diffusion process that we aim to learn the Gaussian transition from a Markov chain, the joint distribution formulated as follows:
\begin{equation}
    \small
    p_{\theta}(Y_{0:T}|D) := p(Y_T)\prod_{t=1}^Tp_{\theta}(Y_{t-1}|Y_t,D)\,,
\label{eq:denoiseRankReverseJointDistribution}
\end{equation}
\begin{equation}
    \small
    p_{\theta}(Y_{t-1}|Y_t,D) := \mathcal{N}(Y_{t-1};\mu_{\theta}(Y_t, t, D), {\scriptstyle\sum}_{\theta}(Y_t, t, D))\,.
\label{eq:denoiseRankReverseJointDistribution2}
\end{equation}
For the forward process, we fixed the approximate posterior $q(Y_{1:T}|Y_0, D)$ to a Markov chain that gradually adds Gaussian noise into the labels:
\begin{equation}
    \small
    q(Y_{1:T}|Y_0) := \prod_{t=1}^Tq(Y_{t}|Y_{t-1})\,,
\label{eq:denoiseRankForwardJointDistribution}
\end{equation}
\begin{equation}
    \small
    q(Y_{t}|Y_{t-1}) := \mathcal{N}(Y_{t};\sqrt{1-\beta_t}Y_{t-1},  \beta_t \textbf{I})\,,
    \label{eq:denoiseRankForwardJointDistribution2}
\end{equation}
\noindent where $\beta_1,...,\beta_T$ is scheduled to control the process of noising data. Let $\alpha_{t} := 1 - \beta_{t}$ and $\overline{\alpha}_{t} := {\prod}_{s=1}^t \alpha_{s}$, then the posterior can be formulated as:
\begin{equation}
    \small
    q( Y_{t} \mid  Y_{0} )=\mathcal{N}(Y_{t};  \sqrt{\overline{\alpha} _{t} } Y_{0}, (1-\overline{\alpha} _{t})\textbf{I} ) \,.
    \label{eq:denoiseRankforwardprioronestep}
\end{equation}

Applying Bayesian theory, the prior probability can be formulated as:
\begin{equation}
    \small
    q( Y_{t-1} \mid  Y_{t}, Y_0 )=\mathcal{N}(Y_{t-1};  \tilde{\mu}(Y_t,Y_0), \tilde{\beta}_tI )\,,
    \label{eq:denoiseRankreverseposteriorBayes}
\end{equation}
\noindent where $\tilde{\beta}_t$ and $\tilde{\mu}(Y_t,Y_0)$ are, respectively, as follows:
\begin{equation}
    \small
    \tilde{\beta}_t := \frac{1-\overline{\alpha}_{t-1}}{1-\overline{\alpha}_{t}}\beta_t\,,
    \label{eq:reverseposteriorbeta}
\end{equation}
\begin{equation}
    \small
    \tilde{\mu}(Y_t,Y_0) := \frac{\sqrt{\overline{\alpha}_{t-1}}\beta_t}{1-\overline{\alpha}_t}Y_0 + \frac{\sqrt{\alpha_t}(1-\overline{\alpha}_{t-1})}{1-\overline{\alpha}_t}Y_t\,.
    \label{eq:reverseposteriormute}
\end{equation}

%
\noindent We then train the model to optimize the variational lower bound (VLB) on negative log likelihood:
\begin{equation}
    \small
\begin{split}
    \mathcal{L} = & \mathbb{E}_q\left[-\log~p(Y_T)-\sum_{t\geq1}\log\frac{p_\theta(Y_{t-1}|Y_t, D)}{q(Y_t|Y_{t-1})}\right]\\ = & \mathbb{E}_q\left[-\log\frac{p_\theta(Y_{0:T}|D)}{q(Y_{1:T}|Y_0)}\right] \geq \mathbb{E}[-~\log~p_\theta(Y_0|D)]\,.
    \label{eq:negativeloglikelihood}
\end{split}
\end{equation}

Thus we can optimize the $\mathcal{L}$ with stochastic gradient descent during training. Further more, the VLB above can be rewritten as follows:
\begin{equation}
    \small
\label{eq:negativeloglikelihood}
\begin{split}
\mathcal{L} &= \mathbb{E}_q\Bigg[ 
    \underbrace{D_{\text{KL}}\big(q(Y_T|Y_0) \parallel p(Y_T)\big)}_{\mathcal{L}_T} \\
    &\quad + \sum_{t>1} \underbrace{D_{\text{KL}}\big(q(Y_{t-1}|Y_t) \parallel p_\theta(Y_{t-1}|Y_t,D)\big)}_{\mathcal{L}_{t}} \\
&\quad - \underbrace{\log p_\theta(Y_0|Y_1)}_{\mathcal{L}_0} \Bigg]\,.
\end{split}
\end{equation}

While $\mathcal{L}_T$ does depend on $\theta$, it will become zero when data being carefully noised enough and can be ignored when optimizing; $\mathcal{L}_0$ 
 use to evaluate the reconstruct quality, but need complex calculate; $\mathcal{L}_t$  calculated by the KL-diveergence between posterior and model prediction.Following \citep{ho2020denoising}, we can simplely rewrite the $\mathcal{L}$ and replace by a mean-squared error (MSE) loss as follows:
\begin{equation}
    \mathcal{L} = \mathbb{E}_{t,Y_0,\varepsilon}[\mid\mid \epsilon - \epsilon_{\theta}(D,Y_t,t) \mid\mid^2]\,,
    \label{eq:denoiserankLossfunctionMSEwithEpsilon}
\end{equation}
where $\epsilon$ and $\epsilon_{\theta}(D,Y_t,t)$ are noise injected to $Y_t$ and noise predicted from denoise neural network of DenoiseRank, respectively. According to \citep{nichol2021improved}, we can further predict $Y_0$ via:
\begin{equation}
    Y_0 = \frac{1}{\sqrt{\alpha_t}}\left( Y_t - \frac{\beta_t}{\sqrt{1-\overline{\alpha}_t}}\epsilon \right)\,.
    \label{eq:denoiserankEpsilonPredictY0}
\end{equation}
Thus, the model can be trained to directly predict $Y_0$ and the loss reformulated as:
\begin{equation}
    \mathcal{L} = \mathbb{E}_{t,Y_0,p_{\theta}}[\mid\mid Y_0 - p_{\theta}(D,Y_t,t) \mid\mid^2]\,,
    \label{eq:denoiserankLossfunctionMSEwithEpsilon}
\end{equation}
\noindent where $p_{\theta}(\cdot)$ is the denoising model we aim to train. That means we can also predict $Y_0$ by the other losses of LTR. In this paper, we do ablation experiments on popular LTR losses to investigate the best performance of our models; see Appendix \ref{sec:Lossfunction}.

\subsection{Diffusion and Training}
In the diffusion process, the main task is to add noise to the labels. Specifically, given time steps $t$ and a noise scheduler $\beta$, the noised labels $Y_t$ are obtained. Our goal is to train an approximate model $p_{\theta}(\cdot)$ that is able to predict the true labels $y_0$ from the noisy labels $y_t$ at $t$ timesteps. We experimented with a variety of noise schedulers, see Appendix \ref{sec:Noisescheduler}. We set the maximum time step T to 1000, and the ablation experiments are referenced in Appendix \ref{sec:Maximumdiffusiontimesteps}.


\paragraph{Training.} We perform hundreds of epocs of training for DenoiseRank as follows. First, the timestep $t$ is randomly sampled for each query list. Based on the time step $t$ and the noise scheduler $\beta$, label vector $Y_0$ will gradually become Gaussian noise $Y_T$ at $T$ (Eq.\ref{eq:denoiseRankforwardprioronestep}). Second, $Y_t$ and document list $D$ are fed into the model $p_{\theta}(\cdot)$, which will output the denoised labels $\hat{Y_0}$. Finally, the loss function is calculated and the model parameters are adjusted. The above process is repeated until the target epoc and the model converge; see Algorithm  \ref{alg:diffusion-training}. Our goal is to make $\hat{Y_0}$ accurately equal to $Y_0$ through extensive training. Training process can be briefly formulated as:
\begin{equation}
\small
    Y_t \leftarrow q(Y_t | Y_0), \space\space \hat{Y}_0 = p_{\theta}(D, Y_t, t)\,.
    \label{eq:denoiserankdiffusion}
\end{equation}

\begin{algorithm}[!t]
\small
\caption{Training}
\label{alg:diffusion-training}
\SetAlgoLined
\DontPrintSemicolon
\KwIn{
    Docs: $D = \{d_1, d_2, \ldots, d_n\}$ \\
    Truth labels: $Y_0 = \{y_1, \ldots, y_n\}$ \\
    Timesteps: $t \in [0, T]$ \\
    Noise schedule: $\beta_t$ \\
    Base model: $p_\theta(\cdot)$ \\
    Training epochs: $K$
}
\KwOut{
    Well-trained Model: $p_{\theta}(\cdot)$
}

\For{epoch $\leftarrow 1$ \KwTo $K$}{
    $Y_t \leftarrow q(Y_t | Y_0)$\tcp*{Eq.(\ref{eq:denoiserankdiffusion})} 
    $\hat{Y}_0 \leftarrow p_\theta(D, Y_t, t)$\tcp*{Prediction} 
    $\mathcal{L} (\hat{Y}_0, Y_0)$\tcp*{Compute loss} 
    $\theta \leftarrow \theta - \eta\nabla_\theta\mathcal{L}$\tcp*{Update} 
}
\Return{$p_\theta$}
\end{algorithm}
%
\subsection{Reverse and Inferencing}

The main task in the reverse process is labels denoising. Starting from a given noisy label $Y_t$ and time step $t$, denoising yields the next time step labels $Y_{t-1}$. The goal of inference is to obtain the predicted labels $\hat{Y}_0$ that can correctly rank the query document $D$. We need to predict $\hat{Y}_0$ at each step and then compute $Y_{t-1}$, see Eq.\ref{eq:denoiseRankReverseJointDistribution2} and Eq.\ref{eq:denoiserankEpsilonPredictY0}.

\paragraph{Inference.} First, the noised labels $Y_t$ is sampled from the Gaussian distribution $\mathcal{N}(0, I)$ in the max timestep $T$. Then, $Y_t$ and the document list $D$ are fed into the model $P_{\theta}$ and the denoised labels $\hat{Y_0}$ are output. According to equation \ref{eq:denoiseRankReverseJointDistribution2} and equation \ref{eq:denoiserankEpsilonPredictY0}, we can approximate $Y_{t-1}$ from $Y_t$ and $\hat{Y_0}$. Finally, repeat the above denoising process until $t=0$, and obtain $y_0$, the predicted relevance labels, see Algorithm \ref{alg:diffusion-inference}. Inferencing process can be brief formulated as:
\begin{equation}
\small
    \hat{Y}_0 = p_{\theta}(D, Y_t, t), ~~ Y_{t-1} \leftarrow q(Y_{t-1} | \hat{Y}_0, Y_t)\,.
    \label{eq:denoiserankreverse}
\end{equation}

\begin{algorithm}[!t]
\small
\caption{Inference} 
\label{alg:diffusion-inference} 
\SetAlgoLined
\DontPrintSemicolon
\KwIn{
    Docs: $D = \{d_1, d_2, \ldots, d_n\}$ \\
    Noised labels: $Y_t \sim \mathcal{N}(0, \mathbf{I})$ \\
    Max Diffusion steps: $T$ \\
    Noise schedule: $\beta_t$  \\
    Trained model: $p_\theta(\cdot)$
}
\KwOut{
    Denoised labels: $\hat{Y}_0$
}
\For{$t \gets T \leftarrow 1$}{  
    $\hat{Y}_0 = p_{\theta}(D, Y_t, t)$ \tcp{Prediction} 
    $q(Y_{t-1} | \hat{Y}_0, Y_t) \to Y_{t-1}$ \tcp{Eq.(\ref{eq:denoiserankreverse})} 
}
\Return{$\hat{Y}_0$} 
\end{algorithm}

\subsection{Discussions}
Our DenoiseRank differs from previous diffusion based generative models in at least the following aspects: (1) Our model is the first one to address T-LTR task by generative diffusion model to accurately rank by fitting the conditional distribution $P(Y|D)$. (2) Previous diffusion models use the U-net as the denoise network (especially in CV areas~\citep{rombach2022high,ho2022video}), but we use a novel network consists of feedforward network and transformer. (3) Compare to the diffusion models in sequence recommendation\citep{li2023diffurec}, we utilise the transformer to calculate context-wise features and denoise labels by the feedforward network.

\section{Experiments}
\label{sec:Experiments}

\subsection{Research Questions}
The remainder of this paper is guided by the following research questions: (1) Does our DenoiseRank outperform state-of-the-art LTR models? (2) How do the hyperparameters and each design choice of the DenoiseRank affect its performance? (3) What about the diversity of the rank result of DenoiseRank compared to other LTR models ?

\subsection{Datasets and Metric}

\paragraph{Datasets.} Experiments were conducted on three famous datasets for LTR, including the Microsoft Web30k \citep{qin2013introducing}, Yahoo! LETOR \citep{chapelle2011yahoo}, and Istella LETOR \citep{dato2016fast}. The queries and documents in these datasets were obtained from real search engines. 
Each dataset contains a large number of query-documents. In addition, labels are annotated on a 5-level scale from 0 to 4 (the most relevant). Each document is represented by multidimensional features such as BM25 score of the page section. We use pre-partitioned train/test data from each dataset for training and testing.

\paragraph{Metrics.} Normalised Discounted Cumulative Gain (NDCG) is used for evaluation purposes. We report the values of the metric at positions 1, 5, and 10, i.e., NDCG@1, NDCG@5, and NDCG@10. It is worth noting that we use NDCG@10 as the evaluation criterion to select the best model. Experimental results against other metrics are shown in Appendix~\ref{sec:Othermetric}.

\subsection{Comparison Models} We evaluate DenoiseRank against the advanced and recent baselines; 
see Table \ref{tab:main-results}. There are two categories of baselines, tree-based and neural-based. 

\paragraph{Tree-based.} $\lambda\mathrm{MART}_{\mathrm{GBM}}$ \citep{ke2017lightgbm} is one of the LambdaMART \citep{wu2010adapting} implementations provided by Microsoft and is one of the best tree-based methods. $\lambda\mathrm{MART}_{\mathrm{RankLib}}$ is part of the RankLib library and is another implementation of LambdaMART~\citep{wu2010adapting}. 

\paragraph{Neural-based.} DLCM~\citep{ai2018learning} employs the RNN network to capture the local ranking context and is trained with an attention-based loss function, which makes it more effective. $\mathrm{SetRank}_{\mathrm{re}}$ \citep{pang2020setrank} is an improved version of SetRank, where documents are reranked before they are entered. NeuralNDCG \citep{pobrotyn2021neuralndcg} is a neural LTR model that addresses the mismatch between optimisation objective and evaluation criterion of traditional model by a novel loss function. DASALC \citep{qin2021neural} is a baseline provided by Google, which is based on the self-attention and multiple optimisation components to outperform even the tree-based models. Rankformer \citep{buyl2023rankformer} is one of the recent baselines, which uses list-wise labels to capture contextual information and also uses an novel implicit feedback component. However, implicit feedback component of Rankformer is not enabled in the experiments as it is not part of our study.

We report evaluation results for $\lambda\mathrm{MART}_{\mathrm{GBM}}$, $\lambda\mathrm{MART}_{\mathrm{RankLib}}$, DLCM, and $\mathrm{Rankformer}$ in the same runtime environment as DenoiseRank. The results for NeuralNDCG, $\mathrm{SetRank}_{\mathrm{re}}$ and DASALC are those reported in their original papers. The DenoiseRank model parameters and training setups are presented in Appendix \ref{sec:Hyperparameters}.

\subsection{Comparison Result}

\begin{table*}
  \centering
  \footnotesize
  \caption{NDCG@K performance comparison on benchmark datasets. Best performance is bolded. 
            \textsuperscript{*} and \textsuperscript{$\dagger$} denote statistically significant 
            improvements over best tree-based and neural models respectively. 
            Last row shows relative difference of DenoiseRank over best comparison models.}
  \label{tab:main-results}
  \begin{tabular}{l c c c c c c c c c}
    \toprule
    Method 
    & \multicolumn{3}{c}{Microsoft Web30K} 
    & \multicolumn{3}{c}{Yahoo!} 
    & \multicolumn{3}{c}{Istella} \\
    \cmidrule(r){2-4} \cmidrule(r){5-7} \cmidrule(r){8-10}
    & @1 & @5 & @10 & @1 & @5 & @10 & @1 & @5 & @10 \\
    \midrule
    $\lambda\mathrm{MART}_{\mathrm{RL}}$       & 45.35  & 44.59  & 46.46  & 68.52  & 70.27  & 74.58  & 65.71  & 61.18  & 65.91 \\
    $\lambda\mathrm{MART}_{\mathrm{GBM}}$       & 50.73  & 49.67  & 51.46  & \textbf{71.90}  & \textbf{74.20}  & 78.01  & \textbf{74.95}  & \textbf{71.20}  & \textbf{76.05} \\
    \midrule
    DLCM            & 46.31  & 45.01  & 46.90  & 67.71  & 69.91  & 74.29  & 65.57  & 61.95  & 66.80 \\
    SetRank$_{re}$  & 45.91  & 45.15  & 46.96  & 68.22  & 70.29  & 74.53  & 67.60  & 63.45  & 68.34 \\  
    NeuralNDCG      & --     & 51.45  & 53.49  & --     & 66.02  & 71.02  & --     & --     & --     \\
    DASALC          & 50.95  & 50.92  & 52.88  & 70.98  & 73.76  & 77.66  & 72.77  & 70.06  & 75.30 \\
    Rankformer      & 49.61  & 49.23  & 51.27  & 70.18  & 73.02  & 77.58  & 68.11  & 68.20  & 75.03 \\
    \midrule
    DenoiseRank     & \textbf{51.87}\textsuperscript{*$\dagger$} & \textbf{52.52}\textsuperscript{*$\dagger$} & \textbf{54.60}\textsuperscript{*$\dagger$} 
                    & 71.37\textsuperscript{$\dagger$} & 74.06\textsuperscript{$\dagger$} & \textbf{78.42}\textsuperscript{*$\dagger$} 
                    & 70.00  & 69.30  & 75.82\textsuperscript{$\dagger$} \\
    \midrule
    \textit{Relative Diff} & +1.8\% & +2.1\% & +2.1\% & $-$0.7\% & $-$1.6\% & +0.5\% & $-$7.1\% & $-$2.7\% & $-$1.4\% \\
    \bottomrule
  \end{tabular}
  
  \medskip
  \footnotesize
  \textit{Note}: $\lambda\mathrm{MART}_{\mathrm{RL}}$ and $\lambda\mathrm{MART}_{\mathrm{GBM}}$ denote Ranklib and GBM versions of LambdaMART.
\end{table*}

The results are shown in Table \ref{tab:main-results}, and we conclude that 1) DenoiseRank achieves better or competitive performance compared to other discriminative models, especially on the Web30k and Yahoo datasets, which proves the effectiveness of our models.  2) DenoiseRank achieves an overall lead over advanced neural LTR baselines. 3) DenoiseRank performs better on the Web30k and Yahoo datasets, and @10 achieves a lead on the Istella dataset compared to other neural LTR models. 4) Compared to advanced tree-based models, DenoiseRank consistently leads in performance on the Web30k dataset, but ahead only at NDCG@10 on the Yahoo dataset and slightly perform worse at NDCG@1, NDCG@5 and every cut-off on the Istella datasets.

Why does DenoiseRank perform best on Web30k, but slightly worse on Yahoo and Istella? Through hypothesis testing, we found that: (1) Web30K contains fewer effective features, and the diffusion-based architecture endows DenoiseRank with robust distributed learning capabilities and uncertainty estimation, making it more resilient to sparse data. (2) Queries in Web30K predominantly fall within medium-to-long spans. The diffusion's noise learning capability makes DenoiseRank particularly adept at handling complex medium-to-long document sequence ranking tasks, though it demands higher training resources. (See Appendix \ref{sec:MainResultAnalyse} for details)

These factors contribute to DenoiseRank's superior performance on Web30K. However, it must be acknowledged that model performance is not determined by a few factors alone. Moving forward, we will conduct deeper analysis and refine DenoiseRank's design to enhance its performance in low-resource environments and short document sequence scenarios.


\subsection{Design Choice and Hyperparameters}

We carry out ablation studies on DenoiseRank and investigate the best design choice of it on different datasets. The configurations of DenoiseRank are in Table \ref{tab:ablation-results}. We further introduce hyperparameters and design choice in Appendix. \ref{sec:Hyperparameters}
 and \ref{sec:AblationStudy}.
\paragraph{Noise schedule.} Noise schedule is the dynamic parameter $\overline{\alpha}_t$ which controls the noise ratio each step during diffusion. We choose 4 types of schedule to evaluate our DenoiseRank, including Linear, TruncatedLinear, Sqrt and Cosine. As shown in Figure~\ref{fig:noiseScheduler} and Table~\ref{tab:noiseSchedulerResult}, we summarize that: (1) Difference schedule affect the performance of our DenoiseRank (2) TruncatedLinear is the better choice because it results in more reliable performance. More introduction is found in Appendix \ref{sec:Noisescheduler}.

\paragraph{Max diffusion timesteps.} Diffusion timesteps $T$ control the speed of noising corresponding labels, noise becomes more subtle as max timestep increases. We choose 5  max timesteps, including 1,000, 800, 600, 400 and 200. The result shown in Figure \ref{fig:maximumDiffusionTimesteps} and Table \ref{tab:maximumDiffusionTimestepsResult} denotes that: (1) The performance is benefitial from carefully nosing after increasing the max diffusion timesteps (2) Training on different datasets need different diffusion timesteps, for example, the best choice is T=600 on istella datasets. Appendix \ref{sec:Maximumdiffusiontimesteps} shows the detailed discussion.

\paragraph{The number of denoise network layers.} Denoise network is an important part of our model to predict ground-truth labels, and its layer count may affect the ranking result of DenoiseRank. Thus we choose 4 types of layers to investigate it, and results are shown in Figure~\ref{fig:denoiseNetLayers} and Table \ref{tab:denoiseLayersResult} denotes that: (1) The number of Layers significantly affects the performance of DenoiseRank (2) the best choice is 2, 4, 4 on Web30K, Yahoo!, Istella datasets respectively. (See Appendix \ref{sec:TheNumberofdenoisenetwroklayers})

\paragraph{Self attentions.} The performance of DenoiseRank is significantly improved as Transformer becomes a part of our model compared to a pure feedforward implementation, as shown in Table~\ref{tab:selfAttentionResult} and Figure~\ref{fig:selfattentions}. The self-attention mechanism makes it passible to recalculate documents feature context-wise, and DenoiseRank can learn about the relation between documents. (See Appendix \ref{sec:SelfAttention})
  
\paragraph{Loss Functions.} In addition to the MSE loss that origin DDPMs employ\citep{ho2020denoising}, we train DenoiseRank with the other 5 famous LTR loss to find to optimal one, including RMSE, RankNet\citep{burges2005learning}, NDCGLoss$_{2++}$\citep{wang2018lambdaloss}, ApproxNDCG\citep{qin2010general,bruch2019revisiting} and ListNet\citep{cao2007learning}. The results are shown in Table \ref{tab:lossFunctionResult} and denotes that: MSE loss perform best in Yahoo! and Istella datasets, while ListNet is the best choice for Web30K. (See Appendix \ref{sec:Lossfunction})

\paragraph{Learning Rate.} We train our model with different learning rates $\in {10^{-1},10^{-2},10^{-3},10^{-4}}$, and user AdamW optimizer. In most situations, $10^{-4}$ is a good choice, while training on the web30k dataset, performance is a little better when learning rate is $10^{-3}$.

\begin{table}[!t]
\centering
\small  
\caption{Recommended configurations of DenoiseRank from ablation study.}  
\label{tab:ablation-results}  
\begin{tabular}{@{}lccc@{}}  
\toprule
\multicolumn{1}{l}{\textbf{Design}} & \textbf{Web30K} & \textbf{Yahoo} & \textbf{Istella} \\
\midrule
Noise Schedule       & TruncL & TruncL & TruncL \\
Max Diffusion Steps  & 1000            & 1000            & 600             \\  
Denoising Layers     & 2               & 4               & 8               \\
Self-Attention       & \checkmark      & \checkmark      & \checkmark      \\  
Loss Function        & ListNet         & MSE             & MSE             \\
\bottomrule
\end{tabular}

\raggedright\footnotesize
\textit{Note}: All configurations use the same base architecture. \checkmark\ denotes the inclusion of self-attention modules. TruncL denotes Truncate Linear schedule.
\end{table}

\subsection{Diversity of Ranking Results} 

As our DenoiseRank is a diffusion based model, it is able to produce diverse ranked lists of documents in response to the same query while may still keep the high standard of the NDCG Performance, compared to traditional LTR models that always produce the same ranked list of documents in response to the same query. 
Such diverse ranked lists of documents allow documents with the same ground-truth labels have the same chances to be ranked in the top-K position in the ranked lists.

In this study, we verify the diversity of the ranking sequences produced by DenoiseRank. In the inference stage, $Y_T$, which is sampled at random from Gaussian noise, introduces uncertainty when making predictions. To address this, we perform multiple inferences for the same query and analyze the ranking diversity. We introduce the RSD@(K,M) metric (see Eq.~\ref{eq:RankingSequenceDiversity}), which denotes the number of different sequences on the top $K$ ranking of the same query among $M$ times inference. (Note that RSD (Ranking Sequence Diversity) is absolutely different to   traditional diversity metrics; see Appendix.~\ref{sec:RankingDiversity}) Using $K\in \{1,5,10,20\}, M=10$, we compare RSD@(K,M) of DenoiseRank to Rankformer, which has a similar architecture to ours but addresses LTR task from a traditional discriminative perspective without uncertainty.

As shown in Figure~\ref{fig:diversityRankingtSNE}, the ranking results inferred by DenoiseRank are diverse, while those inferred by Rankformer remain singleton. Referring to Table \ref{tab:RankingSequenceDiversity}, we find that the ranked lists generated by  DenoiseRank are various in response to the same test query 10 times while the NDCG metric remains excellent. In contrast, Rankformer keeps the same ranked list and the NDCG is also unchanged. However, while ranking diversity, in a few extreme cases, less-related documents may appear at the top, leading to low NDCG@K. It remains one of the future research questions, with optimising sampling noise being a potential approach. More detailed introduce and analysis can be found in the Appendix.~\ref{sec:RankingDiversity}.

\subsection{Efficiency}

DenoiseRank is a diffusion-based model that need to inference by mutiple reverse steps, which may introduced computational overhead. Hence, we investigate testing DenoiseRank's inference time and metrics at different reverse steps, and compared the inference times of the baselines experimentally. All the experiments are conducted on an NVIDIA GeForce RTX 3090 GPU and Intel Xeon CPU Gold 6226R.

As shown in Table \ref{tab:Infer-time-diff-step} and Table \ref{tab:NDCG-diff-step}, inference time of DenoiseRank increases quickly as the reverse step increases while the NDCG performance fluctuates. Therefore, it is unnecessary to use large timesteps that cause computational overhead. We can still achieve excellent NDCG by selecting smaller steps (i.e. 4 reverse steps).

 The comparison experiments was conducted on five baselines, including DenoiseRank, Rankformer, NeuralNDCG, $\lambda\mathrm{MART}_{\mathrm{RankLib}}$ and $\lambda\mathrm{MART}_{\mathrm{GBM}}$. As shown in Table \ref{tab:Infer-time-compare},  the inference time of the DenoiseRank with a smaller reverse step (4 steps) is close to that of NeuralNDCG, but less than the Rankformer. The time processed by the tree-based models is significantly lower than that by the neural network-based models, which is due to their more light-weight model structure and size.

The above experimental results suggest that efficiency is unlikely to become a major optimization bottleneck for DenoiseRank.(We demonstrate efficiency detail in Appendix \ref{sec:effeciency})


\section{Related Work}

Over the last 20 years, 
 Traditional LTR (T-LTR) has usually been studied from the perspective of discriminative methods~\citep{friedman2001greedy,burges2010ranknet,cao2007learning}. These studies can be categorised as tree-based and neural network-based.

Tree-based models show competitive performance~\citep{lucchese2025explainable}, e.g., LambdaMART~\citep{ke2017lightgbm,wu2010adapting}, but poor performance when data is sparse and not easy to be scalable~\citep{qin2021neural}. Other studies advocate the use of neural networks to train LTR models, e.g., RankNet~\citep{burges2005learning}. The advantage of neural networks based models is that they are easy to be scalable, but  are prone to be overfitting, and the feed-forward layer treats documents in isolation and ignore documents' correlation. Some studies introduce attention mechanisms such as RNN and attention 
to LTR and achieve significant performance, e.g., SetRank~\citep{pang2020setrank} and DASALC~\citep{qin2021neural}. However, all these models are discriminative that hard to learn complex distributions. In this work, we introduce generative method to T-LTR, using high-capacity networks and self-attention mechanisms. 

In addition to traditional approachs, some papers address T-LTR by reinforcement learning~\citep{wei2017reinforcement,zhou2020rlirank,xu2020reinforcement,padhye2023deep}. These approachs model ranking as a sequential decision-making process involving multi-step interactions with the environment (documents and queries) to maximise long-term utility. However, this approach incurs substantial computational overhead, with time complexity exhibiting exponential growth compared to traditional methods. This ranking paradigms is different from our DenoiseRank, but it is an intriguing potential study to combine reforcement learning and diffusion models to address LTR~\citep{black2023training}.

Unbiased estimation~\citep{luo2024unbiased} is another LTR studies attracting attention. There have been studies on obtaining implicit feedback and reducing the bias in realistic feedback scores by designing click models, e.g. Rankformer~\citep{buyl2023rankformer}, InfoRank~\citep{jin2024inforank}. Unbiased LTR achieved significant results, but this is far different from our study, which focuses on generative LTR model design. 

 T-LTR studies have used discriminative models, which are also commonly used in classification and regression studies. Generative models, represented by VAE \citep{kingma2013auto}, GAN \citep{goodfellow2014generative}, etc., can model the data distribution \citep{zhou2023deep,liu2021wasserstein} and better solve the problems of data sparsity, overfitting and noise sensitivity, etc. In recent years, there have been researches on the use of generative methods in end-to-end ranking \citep{tamber2023scaling,tamber2025lit,li2024learning}, which employ language models to generate identifiers for ranking. Generative methods provide a new perspective for end-to-end ranking, but it is different from T-LTR, which learn by handcrafted feature. As of now, there is no research on T-LTR using generative models. Diffusion Models \citep{sohl2015deep} have shown great potential (skilled at learning complex distributions) in recent years, with models represented by DDPMs \citep{ho2020denoising,nichol2021improved} being applied to multimodal generation \citep{zhang2025diffclip,song2019generative,ho2022video,song2020improved}. Some studies have applied diffusion models to recommender systems to generate document index one by one \citep{li2023diffurec}. This approach has potential for application to T-LTR, but it's too inefficient, especially when the number of documents to be ranked is large. Overall, this study address T-LTR by diffusion model and lays the foundation for subsequent studies on T-LTR through generative models.


\section{Conclusion and Future Work}

This study aims at addressing the LTR task. Previous studies address the LTR task from a discriminative perspective, do not modeling the data well and ignore the latent relationships among documents. In contrast, to our knowledge we are the first to address the task via a generative model. We propose the novel model, DenoiseRank, which is a diffusion-based LTR model for the task. Specifically, our DenoiseRank noises the relevant labels in the diffusion process and denoises them on the query documents in the reverse process to accurately predict the labels of the documents in response to the input query. Experimental results demonstrate the advantages of our DenoiseRank, including excellent retrieval performance and diversity of ranked lists. We also propose a new evaluation metric for the  performance of the generative model in terms of the diversity of the ranked lists. We believe that our current work makes an important contribution to advance research on neural-based LTR models and paves the way for future research into generative models for LTR.

Future work will explore extending the proposed approach to more challenging settings, including low-resource scenarios and short document sequences. We also plan to investigate more efficient diffusion and sampling strategies to reduce computational overhead while preserving ranking quality. Another promising direction is to integrate complementary paradigms such as reinforcement learning or few-shot learning to enhance robustness and generalization. In addition, we aim to incorporate unbiased and fairness-aware learning objectives to mitigate potential ranking bias. Finally, we will study practical deployment issues, such as stability under distribution shifts and scalability to large-scale real-world systems.

\clearpage
\clearpage

\section*{Limitations}
While DenoiseRank has demonstrated effectiveness in T-LTR tasks, it still has some limitations as follows:

\begin{itemize}
    \item \textbf{Requires substantial data:} DenoiseRank requires substantial data for label distribution generation; with insufficient data, it may underperform
    \item \textbf{Risks flawed outcomes:} While achieving diverse rankings and high average metrics, DenoiseRank risks flawed outcomes in extreme cases.
\end{itemize}

Future work may further explore generative-based LTR with different models, and combine them with few-show learning (FSL) study. Nevertheless, DenoiseRank still offers a novel generative paradigm for traditional LTR, which holds particular significance for future LTR research.

\section*{Ethics Statement}
\label{sec:ethics}
This study aims to construct a diffusion-based LTR model from a generative perspective, enhancing the model's ability to learn complex label distributions, estimate uncertainty, and perform diverse ranking. It provides novel paradigms for traditional LTR while improving performance.

DenoiseRank is purely a research project, utilizing only publicly available benchmark datasets. These datasets originate from real search engine query logs but have been anonymized to exclude any Personally Identifiable Information (PII). Consequently, this research does not involve the collection or use of personal privacy data.

Ranking systems may impact society in practical applications, such as creating information silos or algorithmic bias. Our model, DenoiseRank, has been demonstrated to generate diverse ranking results, which can help mitigate these issues. However, flawed ranking results may still occur in extreme cases. We recommend conducting thorough fairness evaluations and risk assessments before actual deployment.

To promote academic transparency, we have provided the source code repository address in the abstract. This research did not utilize any copyrighted or unauthorized content for training. All authors have reviewed and approved the final version of this paper and declare no potential conflicts of interest

\bibliography{main}

\appendix

\section{Motivation of Adopting Diffusion Models}
\label{sec:Motivation}

\subsection{Weakness of traditional LTR algorithms} 
\begin{itemize}
    \item Given an input query, existing LTR algorithms tend to produce a ranked list of documents that are consistent over time and lack of diversity.
    \item The implicit feedbacks (labels) from users may include a lot of noise. For example, a label of 0 may not mean 'irrelevant', but rather 'no clicks'. Traditional LTR algorithms are restrictive in uncertainty estimation and are sensitive to noise.
    \item The distribution of user feedback is complex and non-linear. Traditional LTR models, which rank from a discriminative perspective, tend to make single-point estimates but not distribution.
\end{itemize}
\subsection{Why not other generative models for LTR} 
\begin{itemize}
    \item GANs, as one category of famous generative models, suffer from mode collapse and training instability issue, which causes them to fall into a local optimum.
    \item VAE, as one category of famous generative models, suffer from posterior collapse and the limited representation capacity issue, restrict the model to rank accuracy.
    \item Autoregressive, as one category of famous generative models, can rank items by predicting the next item, but suffers from error accumulation and slow generation issue.
\end{itemize}
\subsection{Advantages of diffusion models when are integrated into LTR} 
\begin{itemize}
    \item The diffusion model starts the inference process with random noise and perturbs the results during denoising. This provides a diverse ranking results for the LTR task.
    \item In reverse process, diffusion model starts from a random noise, it can be seen as uncertainty about user feedback, and the subsequent denoising progressively estimates the user's intentions.
    \item Diffusion models excel at modelling complex, non-linear mapping relationships and have the ability to generate complex distributions and learn conditional probability distributions, which is desperately needed for learning to rank.
    \item Diffusion model also has the advantage of being stable for training and theoretically rigorous.
\end{itemize}

\section{Denoise Neural Network in DenoiseRank}
\label{sec:DenoiseRankModelArchitecture}

As shown in Figure \ref{fig:DenoiseRank_Diffusion}, Our model contains two components, the Transformer encoder and the feedforward network. In recent years, neural LTR models employ Transformer encoder(self-attention mechanism) and achieve significant performance advances. Since the self-attention mechanism can model the query documents context-aware, we make it as a part of our models. We choose transformer encoder as the basic network, formulated as follows:
\begin{equation}
\small
    \mathbf{H} = \mathbf{E}(\mathbf{d_1,d_2,....,d_n})\,,
    \label{eq:denoiserankTransformerEncoder}
\end{equation}
\noindent where $\mathbf{d_1,d_2,....,d_n}$ are embeddings of all documents in response to a single query, $\mathbf{H}$ are context-wise features of documents calculated by the transformer encoder $\mathbf{E}(\cdot)$. We use a standard transformer encoder architecture, where input document features are computed with advanced self-attention \citep{vaswani2017attention}, followed by feed-forward networks and activation functions, and finally layer normalisation and dropout. In order to increase the model capacity and get the optimal performance, we try multi-head attention and multi-blocks architecture in training, $heads \in \{1,2,4,8\}$ and $blocks \in \{3,4,5,6\}$. Unlike the CV diffusion model, the denoising network is a feed-forward network instead of a U-net, formulated as:
\begin{equation}
\small
    \hat{Y}_0 = \text{FFN}(\mathbf{H}, Y_t, t)\,,
    \label{eq:denoiserankFeedforword}
\end{equation}
\noindent where t is the time step, $Y_t$ is a corresponding noised label at $t$. $\text{FFN}(\cdot)$ is a feedforward network with a multi-tiered architecture \citep{han2022card}. As shown in Figure \ref{fig:DenoiseRank_Diffusion} on the right, firstly, we input $Y_{t,i}$ and $\mathbf{H}_i)$ into the first layer of the network after concatenating them to obtain the output $\mathbf{h}^{(l)}=\mathrm{layer}(\mathbf{H}_i,Y_{t,i})$, where $\mathbf{H}_i$ denotes the $i$th document feature vector and $Y_{t,i}$ denotes the corresponding $i$th label. In each denoising layer, it goes through a linear layer, an activation function and a dropout layer respectively, then we do the embedding calculation to get $t_{\mathrm{emb}}$ for timestep $t$. Finally, $\mathbf{h}^{(l)}$ is multiplied by $t_{emb}$ to get the current output.
In our experiments, we try multiple denoise layer architectures with the number of layers n $\in \{2,4,6,8\}$. The first layer formulated as:
\begin{equation}
\small
    \mathbf{h}^{(1)} =\text{Dropout}(\sigma_{\text{sp}}(\text{Linear}(\mathbf{H}_i, Y_{t,i})))\,,
    \label{eq:denoiserankFeedforwordLayerfirst}
\end{equation}
\begin{equation}
\small
    \mathbf{m}^{(1)} = \mathbf{h}^{(1)} \odot t_{emb}\,,
    \label{eq:denoiserankFeedforwordLayerfirst2}
\end{equation}
\noindent middle layer formulated as:

\begin{equation}
\small
    \mathbf{h}^{(j)} =\text{Dropout}(\sigma_{\text{sp}}(\text{Linear}(\mathbf{m}^{(j-1)})))\,,
    \label{eq:denoiserankFeedforwordLayermid}
\end{equation}
\begin{equation}
\small
    \mathbf{m}^{(j)} = \mathbf{h}^{(j)} \odot t_{\mathrm{emb}} \,,
    \label{eq:denoiserankFeedforwordLayermid2}
\end{equation}
\noindent where $j$ denotes the $j$-th denoise layer of $\text{FFN}(\cdot)$, $j \in (1, n)$. The output layer is formulated as:
\begin{equation}
\small
    \mathbf{h}^{(n)} = \text{Dropout}(\sigma_{\text{sp}}(\text{Linear}(\mathbf{m}^{(n-1)})))\,,
    \label{eq:denoiserankFeedforwordLayerOutput}
\end{equation}
\noindent where $\sigma_{\text{sp}}$ is the softplus activation function, $\mathbf{m}^{(j-1)}$ is the output of the previous layer. The dimension of the last layer $\mathbf{h}^{(n)}$ is 5, each dimension corresponds to the weight of the relevance label, and the final weighted sum computes the predicted label $\hat{Y}_0$, where $\hat{Y}_0 \in [0,4]$ and is denoted as:
\begin{equation}
\small
    \hat{Y}_0 = \mathbf{h}^{(n)} \odot (0,1,2,3,4)\,.
    \label{eq:denoiserankFeedforwordLayerWeightCal}
\end{equation}

\section{Hyperparameters}
\label{sec:Hyperparameters}

\paragraph{DenoiseRank hyperparameters} We set the hyperparameters of the model, including: dropout $\in \{0.1,0.2,0.3,0.4,0.5,0.6,0.7,0.8\}$, denoise net hiddenSize $\in \{64,128,256,512\}$ for linearLayer, denoise Layers $\in \{2,4,6,8\}$, transformer blocks $\in \{3,4,5,6\}$, self-attention heads $\in \{1,2,4,5,8\}$. In the diffusion configuration, noise schedule $\in \{TruncatedLinear, Linear, Cosine, Sqrt\}$, max diffusion timesteps $\in \{100, 200, 300, 400, 500, 600, 700, 800, 900, 1000\}$. We have experimented and tuned different datasets and the results can be found in Tables \ref{tab:maximumDiffusionTimestepsResult} , \ref{tab:noiseSchedulerResult}. We introduce the experiments result in the following paragraph and Appendix.~\ref{sec:AblationStudy}.

\begin{figure*}
  \centering
  \includegraphics[width=1.0\textwidth]{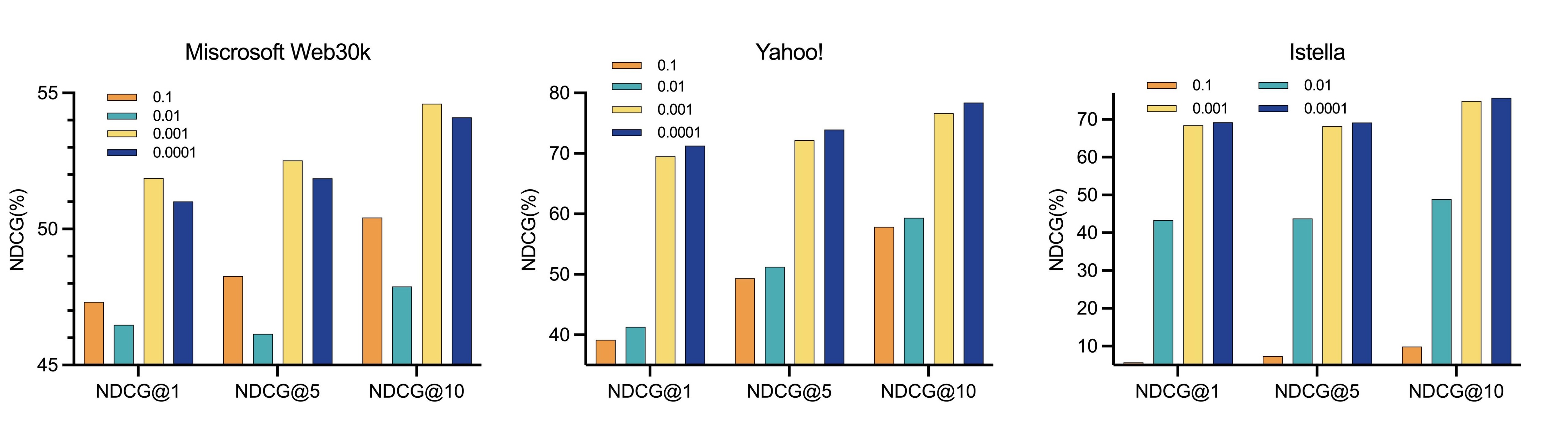}
  \caption{NDCG@K of DenoiseRank with different learning rates on Miscrosoft Web30k, Yahoo! and Istella datasets.}
  \label{fig:learningRateResult}
\end{figure*}

\paragraph{DenoiseRank training settings} First, we use the AdamW optimiser and set LearningRate $\in \{0.1,0.01,0.001,0.0001\}$ and batchsize=128. Second, the training epoc is set to 200. For training, we evaluate every 10 epochs on the validate dataset using NDCG@10 as a benchmark. Store the optimal model and the results evaluated by the metrics. Finally, we run the model on a single NVIDIA GeForce RTX 3090. The performance with different learning rate is shown in Figure~\ref{fig:learningRateResult} and Table~\ref{tab:learningRateResult}. We find that:

\begin{enumerate}
    \item A learning rate of $10^{-3}$ is optimal for training DenoiseRank on the MS Web30K dataset, with $10^{-4}$ being the next best option.
    \item For the Yahoo! and Istella datasets, $10^{-4}$ is the better learning rate with which to train DenoiseRank; $10^{-3}$ provides an approximate result.
    \item In most situations, learning rates of $10^{-1}$ and $10^{-2}$ result in poor performance, which suggests that our DenoiseRank needs subtle optimisation.
\end{enumerate}

\begin{table*}[ht]
  \small
  \centering
  \caption{NDCG@K performance of DenoiseRank with different learning rates on Microsoft Web30K, Yahoo!, and Istella datasets. Best performance per column in bold.}
  \label{tab:learningRateResult}
  \begin{tabular}{@{}lrrrrrrrrr@{}}
    \toprule
    \multirow{2}{*}{LR} & \multicolumn{3}{c}{Web30K} & \multicolumn{3}{c}{Yahoo!} & \multicolumn{3}{c}{Istella} \\
    \cmidrule(r){2-4} \cmidrule(r){5-7} \cmidrule(r){8-10}
    & @1 & @5 & @10 & @1 & @5 & @10 & @1 & @5 & @10 \\
    \midrule
    0.1    & 47.32 & 48.27 & 50.42 & 39.17 & 49.34 & 57.84 & 5.70 & 7.38 & 9.87 \\
    0.01   & 46.48 & 46.15 & 47.89 & 41.31 & 51.24 & 59.34 & 43.39 & 43.80 & 48.89 \\
    0.001  & \textbf{51.87} & \textbf{52.52} & \textbf{54.60} & 69.51 & 72.18 & 76.65 & 68.42 & 68.20 & 74.85 \\
    0.0001 & 51.01 & 51.86 & 54.10 & \textbf{71.27} & \textbf{73.96} & \textbf{78.40} & \textbf{69.14} & \textbf{69.09} & \textbf{75.63} \\
    \bottomrule
  \end{tabular}
\end{table*}

\paragraph{Convergence} DenoiseRank is a new LTR model consider the task from generative perspective, combine with Diffusion model, which need a lot of timesteps in diffusion and reverse process. Thus we investigate the convergence speed on training process in experiments on the runtime enviroment we mention above. We alse compare our model to Rankformer, which has the similiar model architecture. We use the best hyperparameter and design choice of them and which can make a best ranking performence.

\begin{figure*}[t]
  \includegraphics[width=0.48\linewidth]{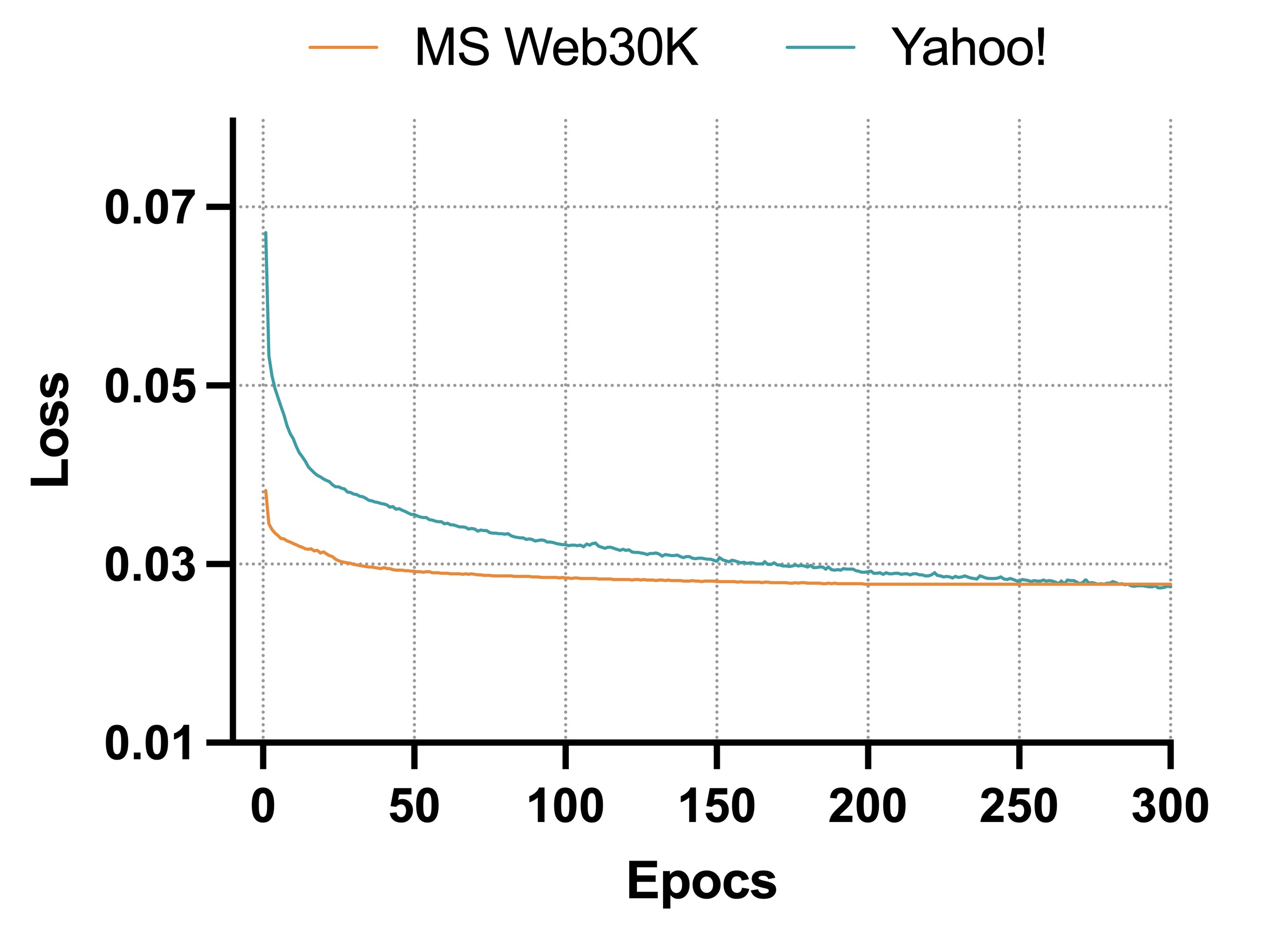} \hfill
  \includegraphics[width=0.48\linewidth]{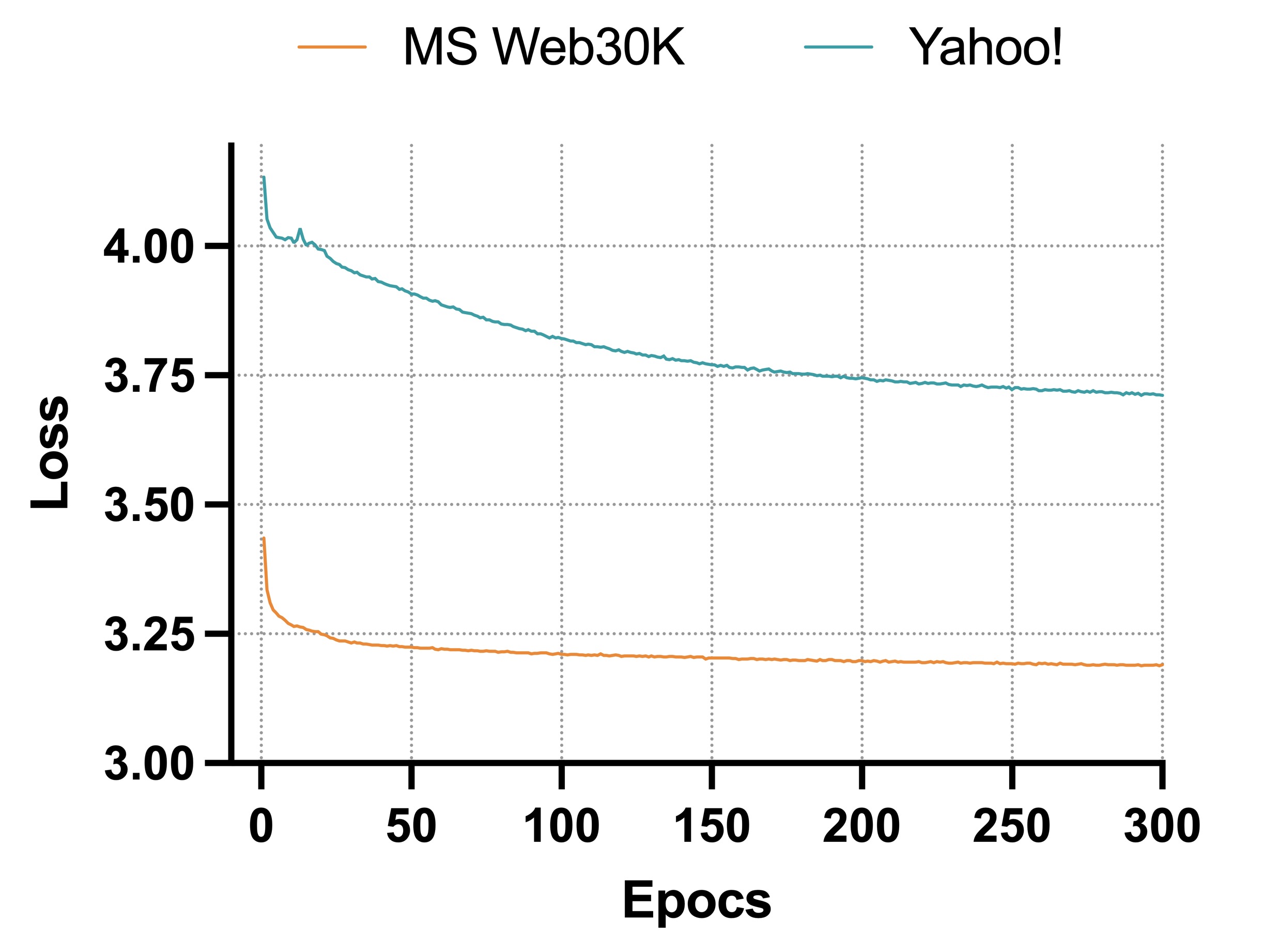}
  \caption {Curve of training loss of DenoiseRank(Left) and Rankformer(Right) on MS Web30K and Yahoo! datasets among 300 epocs.}
  \label{fig:Curveoftraningloss}
\end{figure*}

As shown in Figure.~\ref{fig:Curveoftraningloss}, we summarize that:
\begin{enumerate}
    \item On the MS Web30K datasets, both DenoiseRank and Rankformer can converge after 50 epocs of training.
    \item On the Yahoo! datasets, DenoiseRank converge after 130 epocs, while rankformer is more slow and coverage after 200 epoc.
    \item We speculate it is because: first, documents in Yahoo! have higher dimension of feature (700 dimensions per document) than those in MS Web30K (136 dimensions per document), so model need more epoc to fit them; second, our DenoiseRank address LTR task from generative perspective and comine with Diffusion model, it can fit high dimensional feature more effective
\end{enumerate}

\section{Main Result Analyse}
\label{sec:MainResultAnalyse}
As the main comparison result shown in Table \ref{tab:main-results}, DenoiseRank perform best on Web30k, but slightly worse on Yahoo and Istella. We have proposed two hypotheses: (1) DenoiseRank demonstrates greater robustness for sparse data (with few effective features). (2) DenoiseRank performs better on medium-to-long sequences and requires sufficient training data.

In order to verify the above hypothesis, we analyse number of effective features and query-documents length distribution of different datasets, and construct experiments to evaluate models performance.

\subsection{Number of effective features}
\textbf{Experimental Design:}\\
\textbf{First,} we calculated the average number of effective features(NOEF) across the three datasets to understand their feature characteristics. \textbf{Second,} we uniformly split the train.txt files of the three benchmarks into six versions of training data based on document sequence length(See Table \ref{tab:effectiveFeatureSegment}). Three versions contain high-feature-count documents, while the other three contain low-feature-count documents. Subsequently, we tested each model on the original test.txt file, comparing the test results from high-feature-count and low-feature-count versions (calculating the mean and variance for each of the three versions).

\begin{figure*}
  \centering
  \includegraphics[width=1.0\textwidth]{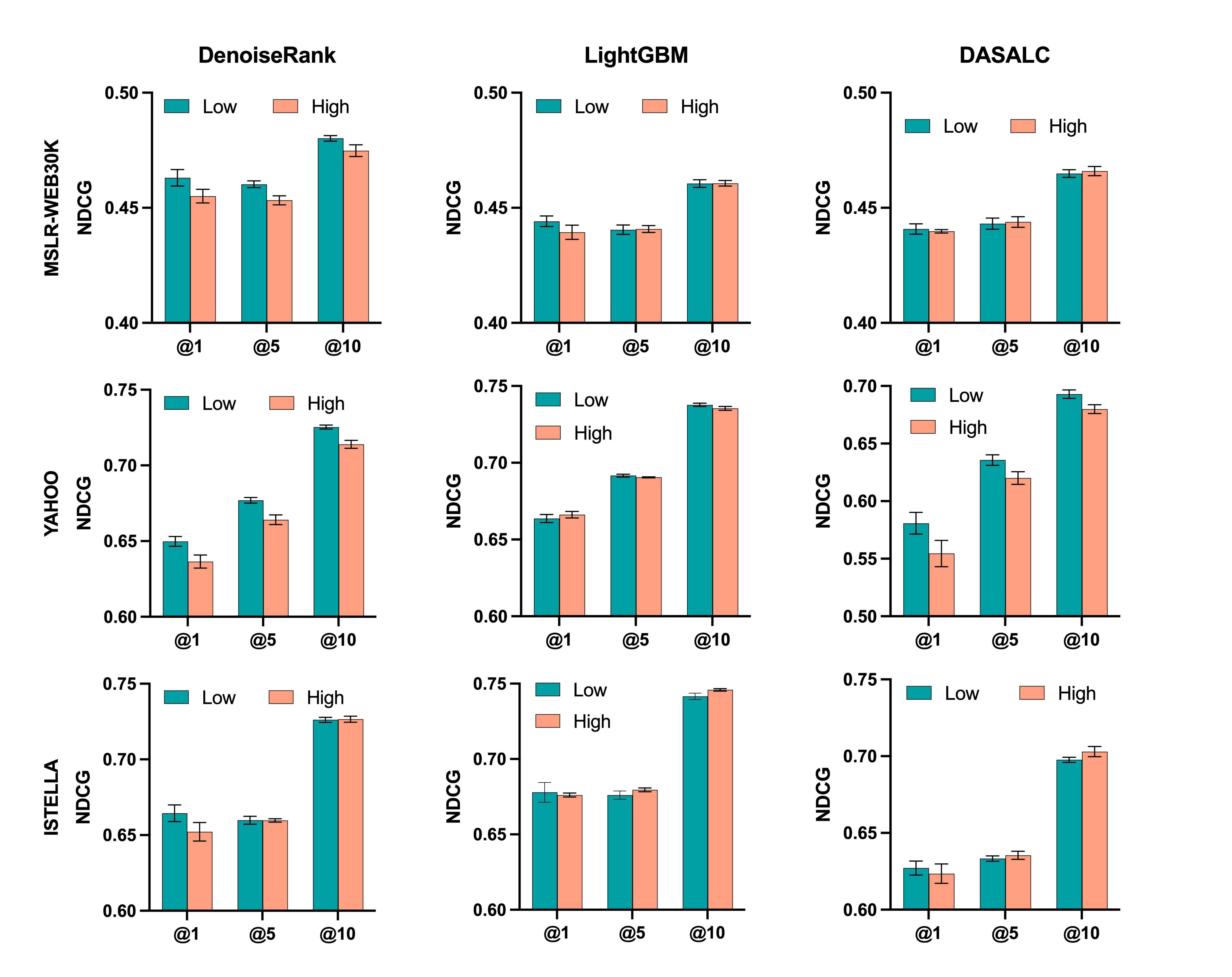}
  \caption{Control group experimental results for effective feature counts, NDCG@K performance produce by different models on Miscrosoft Web30k, Yahoo! and Istella datasets. \textbf{Low} means lower effective feature training resources, \textbf{High} means higher effective feature training resources.}
  \label{fig:FeatureDensityNDCG}
\end{figure*}

\noindent
\textbf{Experimental Results:}\\
As the result shown in  Table \ref{tab:featureDensity} and Figure \ref{fig:FeatureDensityNDCG}:\\
1. On train.txt, YAHOO's NOEF is 224 , ISTLLA's NOEF is  115 , and WEB30K's NOEF is 85. Effective feature count ranking: YAHOO > ISTLLA > WEB30K\\
2. On Web30K, DenoiseRank performs excellently in versions with fewer effective features, while the other two models show little difference.\\
3. On YAHOO, DenoiseRank and DASALC perform excellently in versions with fewer effective features, while GBM shows little difference.\\
4. On ISTELLA, DenoiseRank performs slightly better in versions with fewer effective features but overall performs poorly.

These results suggest that DenoiseRank, being diffusion-based, demonstrates superior learning capabilities for distributions and robustness on sparse features compared to other models. Consequently, it exhibits advantages on WEB30K (with fewer effective features) but shows limited or even inferior performance on YAHOO and ISTELLA (with more effective features). However, feature sparsity is not the sole influencing factor; other contributing factors warrant further exploration.

\begin{table*}[ht]
  \small
  \centering
  \caption{Statistics of different dataset, including queries number, docments number and the average number of effective feature(NOEF).}
  \label{tab:featureDensity}
  \begin{tabular}{@{}lrrrrrrrrr@{}}
    \toprule
    \multirow{2}{*}{Subset} & \multicolumn{3}{c}{Web30K} & \multicolumn{3}{c}{Yahoo!} & \multicolumn{3}{c}{Istella} \\
    \cmidrule(r){2-4} \cmidrule(r){5-7} \cmidrule(r){8-10}
    & Queries & Docs & NOEF & Queries & Docs & NOEF & Queries & Docs & NOEF \\
    \midrule
    train    & 18919 & 2.27M & 85.2 & 19944 & 0.47M & 222.7 & 19245 & 2.04M & 115.6 \\
    vali   & 6306 & 0.75M & 85.4 & 2994 & 0.07M & 223.2 & 7211 & 0.68M & 114.8 \\
    test  & 6306 & 0.75M & 85.3 & 6983 & 0.17M & 222.5 & 6562 & 0.68M & 115.5 \\
    total & 31531 & 3.77M & 85.2 & 29921 & 0.71M & 222.8 & 33018 & 3.41M & 115.4 \\
    \bottomrule
  \end{tabular}

  \medskip
  \footnotesize
  \noindent
  \textit{Note}: M means millions.
\end{table*}

\begin{table*}[ht]
  \small
  \centering
  \caption{Split result of the train.txt files of the three benchmark( Microsoft Web30K, Yahoo!, and Istella). \textbf{Low} means lower effective feature training resources, \textbf{High} means higher effective feature training resources. The numerical representation average number of effective features(NOEF)}
  \label{tab:effectiveFeatureSegment}
  \begin{tabular}{@{}crrrrrr@{}}
    \toprule
    \multirow{2}{*}{Dataset} & \multicolumn{3}{c}{Low} & \multicolumn{3}{c}{High} \\
    \cmidrule(r){2-4} \cmidrule(r){5-7} 
    & S1 & S2 & S3 & S4 & S5 & S6  \\
    \midrule
    Web30K & 76.33 & 76.35 & 76.66 & 94.17 & 94.18 & 94.02  \\
    YAHOO  & 191.38 & 191.10 & 190.61 & 255.36 & 254.31 & 254.52  \\
    ISTELLA  & 109.63 & 109.65 & 109.63 & 121.53 & 121.51 & 121.44  \\
    \bottomrule
  \end{tabular}
\end{table*}

\subsection{Query-Document Length}

\textbf{Experimental Design: }\\
\textbf{First,} we calculated the query-document length distributions across three datasets to understand their length characteristics. \textbf{Second,} we trained models on the training sets of three benchmarks, then analyzed model performance across different query-document length intervals (one interval per 10 lengths).

\begin{figure*}
  \centering
  \includegraphics[width=1.0\textwidth]{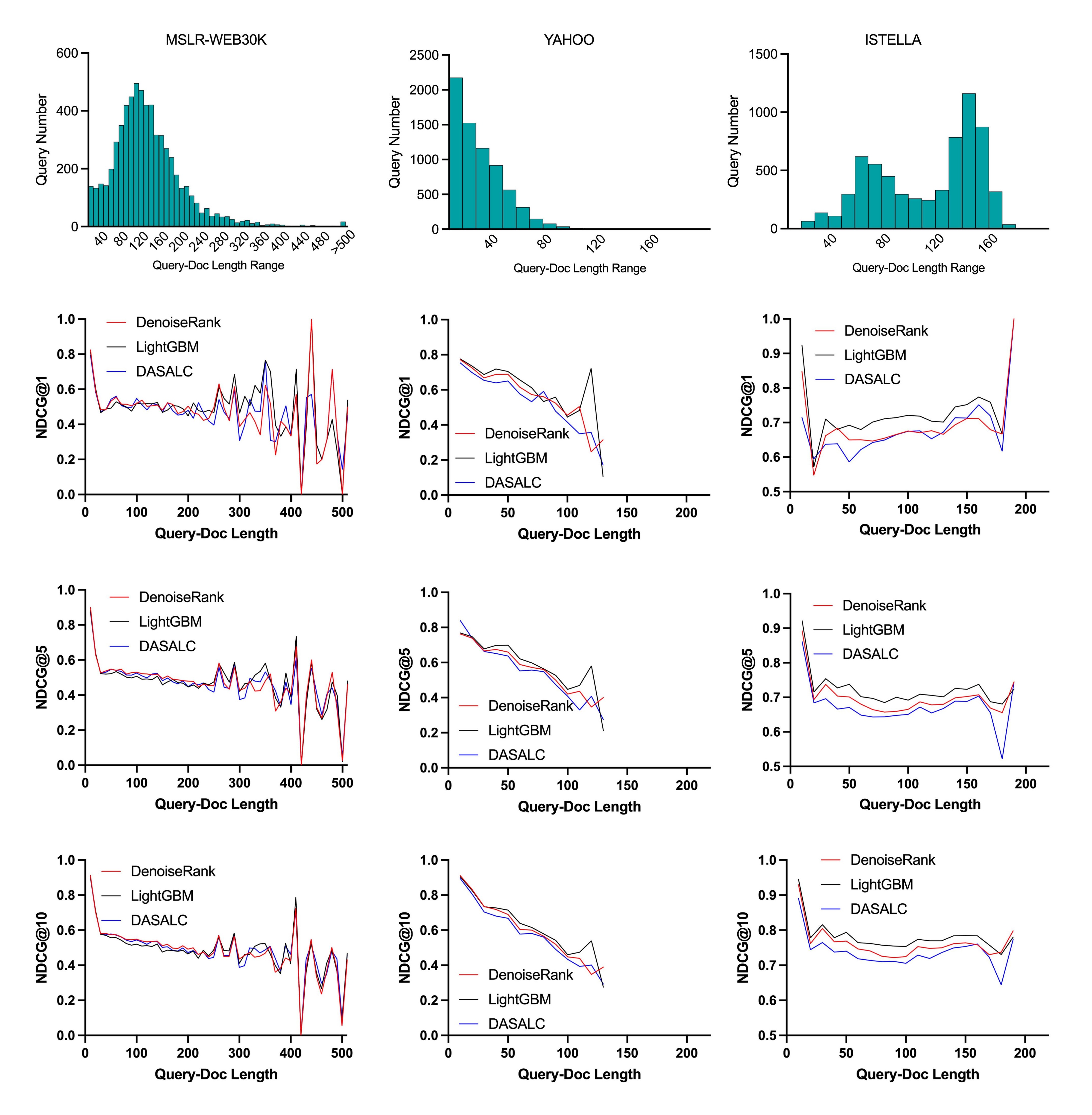}
  \caption{Statistical and Experimental Results of Query-Document Length analysis. \textbf{First row of figure:} Query-Documents length distributions. \textbf{Subsequent row of figure:} NDCG@K of models(DenoiseRank, LightGBM, DASALC) under different length on Miscrosoft Web30k, Yahoo! and Istella datasets.}
  \label{fig:LenDistNDCG}
\end{figure*}

\noindent
\textbf{Experimental Results:}\\
As the result are shown in Figure \ref{fig:LenDistNDCG}: \\
1. the length of query-document in Web30K exhibit a central tendency around 110, following a normal-like distribution, and display characteristics of a long-tail distribution (actually the max length is nearly 1300). In contrast, query-document length in Istella presents a hump distribution (max length < 190), and those in Yahoo gradually decreases between 1 and 120 (max length < 140). \\
2. On the WEB30K dataset, DenoiseRank outperformed the other two models in the medium-length range (50–250), while the difference was negligible in the long-tail range (>250).\\
3. On the YAHOO dataset, DenoiseRank performed similarly to LightGBM in the short range (<50) but underperformed compared to LightGBM in the medium-long range (>50), while DASALC showed poorer performance.\\
4. On the ISTELLA dataset, DenoiseRank underperforms LightGBM across most intervals and slightly trails DASALC in certain ranges (130–160), particularly when @K is reduced.

Experimental results indicate that DenoiseRank achieves optimal performance on medium-to-long intervals with sufficient training resources (e.g., its performance on Web30K), while its advantage is negligible on short intervals even with ample resources (e.g., its performance on YAHOO). However, these findings also remind us that length is not the sole influencing factor. The underlying reason is that diffusion-based and transformer-based DenoiseRank models excel at processing longer document sequences, hence their strong performance on medium-to-long sequences. However, their powerful distribution learning capabilities also demand substantial training resources, leading to suboptimal performance when resources are distributed across shorter sequences (as seen on ISTELLA).

\subsection{Discussion}
In summary, DenoiseRank excels on Web30K but performs slightly worse on Yahoo and ISTELLA because its diffusion-based architecture offers strong distribution fitting capabilities and robustness to sparse features, excelling on medium-to-long document sequences. However, it also demands substantial and concentrated training resources. Frankly, model performance isn't determined by a single factor. Moving forward, we will conduct deeper analysis and refine DenoiseRank's design to enhance its capabilities in low-resource environments and short document sequences.

\clearpage
\section{Ablation Study}
\label{sec:AblationStudy}

We have done ablation studies on DenoiseRank, including maximum diffusion timesteps, noise scheduler, the number of denoise network layers and effectiveness of self-attentions; see Table [\ref{tab:maximumDiffusionTimestepsResult},\ref{tab:noiseSchedulerResult},\ref{tab:denoiseLayersResult}, \ref{tab:selfAttentionResult}].

\subsection{Maximum Diffusion Timesteps}
\label{sec:Maximumdiffusiontimesteps}

\begin{figure*}
  \centering
  \includegraphics[width=0.32\textwidth]{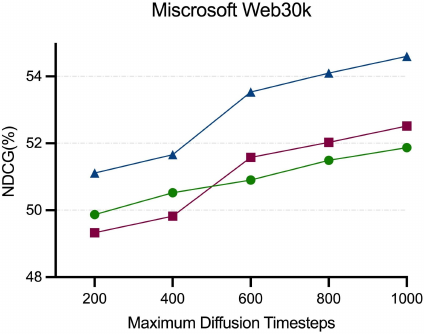}
  \includegraphics[width=0.32\textwidth]{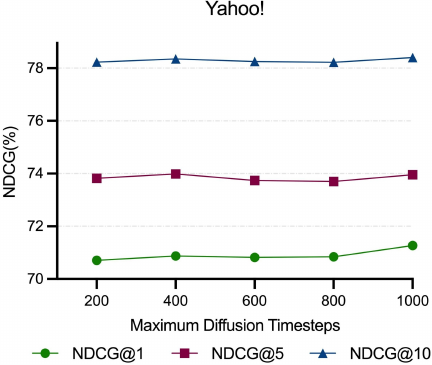}
  \includegraphics[width=0.32\textwidth]{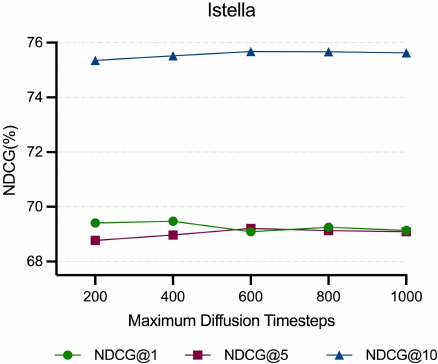}
  \caption{NDCG@K of DenoiseRank at different noise schedule on Miscrosoft Web30k, Yahoo! and Istella datasets.}
  \label{fig:maximumDiffusionTimesteps}
\end{figure*}

In the diffusion process, the maximum diffusion timestep ($T$) refers to how many iterations are required to change from the original sample to the Gaussian noise. The larger the maximum time step, the smaller the sample change per iteration, and conversely, the larger the sample change. The original DDPM uses a maximum time step of $T=1000$ \citep{ho2020denoising}\citep{nichol2021improved}. As different maximum time steps can have an impact on the model performance \citep{li2023diffurec}. We evaluate the model performance in the case of time step $t=[1000,800,600,400,200]$ respectively. As shown in Figure. \ref{fig:maximumDiffusionTimesteps} and Table \ref{tab:maximumDiffusionTimestepsResult}:
\begin{enumerate}
    \item The model performs better as the maximum time step increases, suggesting that slow noise addition is more beneficial for model learning.
    \item The model performance is more dependent on long time steps on the web30k dataset.
    \item The performance of the model is not always optimal for long time steps. The model performs optimally on the Istella dataset at $T=600$. This means that we can reduce the time step appropriately to speed up training and inference.
\end{enumerate}

\begin{table*}[ht]
  \small
  \centering
  \caption{NDCG@K performance with different diffusion timesteps on Microsoft Web30K, Yahoo!, and Istella datasets. Best performance per column in bold.}
  \label{tab:maximumDiffusionTimestepsResult}
  \begin{tabular}{@{}crrrrrrrrr@{}}
    \toprule
    \multirow{2}{*}{Timesteps} & \multicolumn{3}{c}{Web30K} & \multicolumn{3}{c}{Yahoo!} & \multicolumn{3}{c}{Istella} \\
    \cmidrule(r){2-4} \cmidrule(r){5-7} \cmidrule(r){8-10}
    & @1 & @5 & @10 & @1 & @5 & @10 & @1 & @5 & @10 \\
    \midrule
    1000 & \textbf{51.87} & \textbf{52.52} & \textbf{54.60} & \textbf{71.27} & 73.96 & \textbf{78.40} & 69.14 & 69.09 & 75.63 \\
    800  & 51.49 & 52.03 & 54.10 & 70.84 & 73.70 & 78.22 & 69.25 & 69.13 & 75.67 \\
    600  & 50.90 & 51.58 & 53.53 & 70.82 & 73.74 & 78.25 & 69.09 & \textbf{69.21} & \textbf{75.68} \\
    400  & 50.52 & 49.82 & 51.66 & 70.87 & \textbf{73.99} & 78.35 & \textbf{69.47} & 68.97 & 75.52 \\
    200  & 49.87 & 49.33 & 51.11 & 70.71 & 73.82 & 78.23 & 69.41 & 68.77 & 75.35 \\
    \bottomrule
  \end{tabular}
\end{table*}

\subsection{Noise Scheduler}
\label{sec:Noisescheduler}

The noise scheduler is the way in which the $\overline{\alpha}_t$ changes during diffusion, where $\overline{\alpha}_{t} := {\prod}_{s=1}^t \alpha_{s}$, see eq. \ref{eq:denoiseRankForwardJointDistribution2}. The rate of change of $\overline{\alpha}_t$ varies in different noise-adding schemes, e.g., truncated linear has a large change before $\frac{T}{2}$ and a small change after $\frac{T}{2}$, whereas Cosine has a relatively balanced change\citep{li2023diffurec}.

\begin{figure*}
  \centering
  \includegraphics[width=0.32\textwidth]{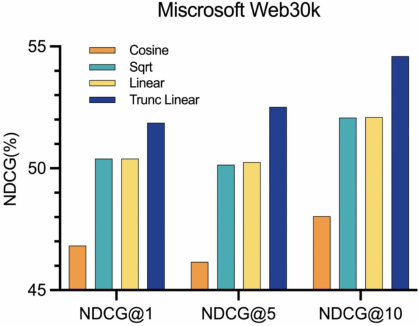}
  \includegraphics[width=0.32\textwidth]{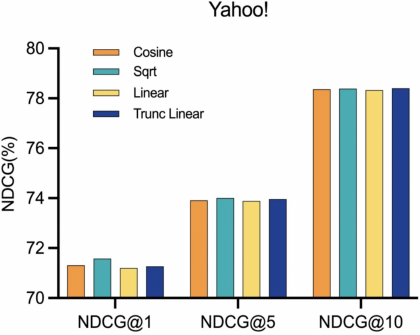}
  \includegraphics[width=0.32\textwidth]{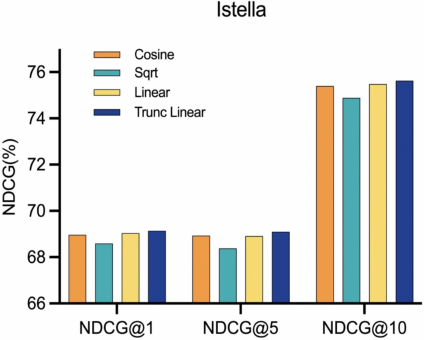}
  \caption{NDCG@K of DenoiseRank at different noise schedule on Miscrosoft Web30k, Yahoo! and Istella datasets.}
  \label{fig:noiseScheduler}
\end{figure*}

In order to evaluate the performance of DenoiseRank under different noise schedules, we try different choices, including Truncated Linear, Linear, Cosine, Sqrt. The results (see Figure.\ref{fig:noiseScheduler} and Table.\ref{tab:noiseSchedulerResult}) show that:
\begin{enumerate}
    \item TruncatedLinear performs better than the other schedules overall, but there is not a big difference.
    \item the performance of the different noise schedules varies greatly on the web30k datasets, i.e. TruncatedLinear $>$ Sqrt $>$ Linear $>$ Cosine.
    \item on the yahoo and istella datasets, there is not much difference in the reliability of the ranking, and on the istella dataset, sqrt even performs slightly better than TruncatedLinear.
\end{enumerate}

\begin{table*}[ht]
  \small
  \centering
  \caption{NDCG@K performance with different noise schedulers on Microsoft Web30K, Yahoo!, and Istella datasets. Best performance per column in bold.}
  \label{tab:noiseSchedulerResult}
  \begin{tabular}{@{}lrrrrrrrrr@{}}
    \toprule
    \multirow{2}{*}{Scheduler} & \multicolumn{3}{c}{Web30K} & \multicolumn{3}{c}{Yahoo!} & \multicolumn{3}{c}{Istella} \\
    \cmidrule(r){2-4} \cmidrule(r){5-7} \cmidrule(r){8-10}
    & @1 & @5 & @10 & @1 & @5 & @10 & @1 & @5 & @10 \\
    \midrule
    TruncatedLinear & \textbf{51.87} & \textbf{52.52} & \textbf{54.60} & 71.27 & 73.96 & \textbf{78.40} & \textbf{69.14} & \textbf{69.09} & \textbf{75.63} \\
    Linear          & 50.40 & 50.25 & 52.10 & 71.20 & 73.89 & 78.33 & 69.04 & 68.91 & 75.48 \\
    Cosine          & 46.83 & 46.16 & 48.04 & 71.31 & 73.91 & 78.36 & 68.59 & 68.38 & 74.89 \\
    Sqrt            & 50.40 & 50.15 & 52.08 & \textbf{71.58} & \textbf{74.00} & 78.39 & 68.97 & 68.93 & 75.40 \\
    \bottomrule
  \end{tabular}
\end{table*}

\subsection{The Number of Denoise Network Layers}
\label{sec:TheNumberofdenoisenetwroklayers}

As shown in Figure. \ref{fig:DenoiseRank_Diffusion} on the right, the denoising network of DenoiseRank is a feed-forward architecture. The input and output layers of the denoising network are required, and the hidden layers in between can be dynamically adjusted \citep{han2022card}. Different hidden layers can affect the performance of the model.
To obtain the optimal model structure, we investigate the effect of different layers of the denoising network. The number of layers includes $[2,4,6,8]$, e.g. $layer=2$ means that only the input and output layers are included and there is no hidden layer.

\begin{figure*}[!t]
  \centering
\includegraphics[width=0.32\textwidth]{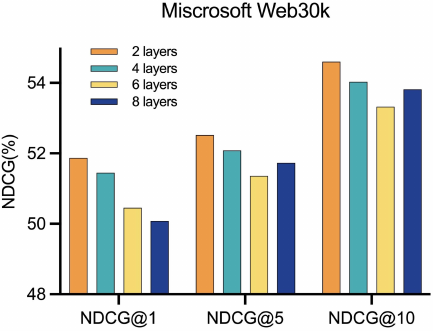}
\includegraphics[width=0.32\textwidth]{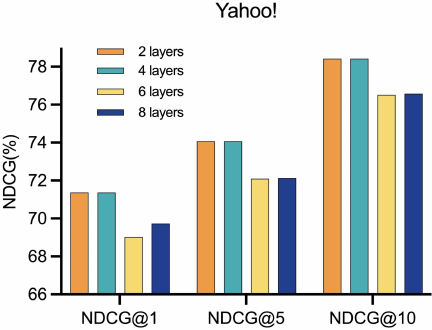}
\includegraphics[width=0.32\textwidth]{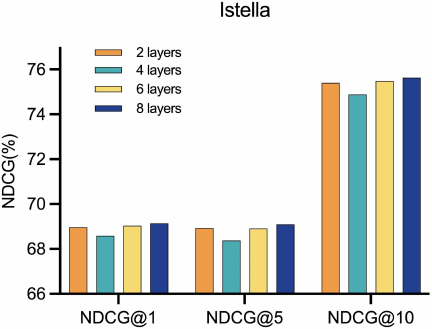}
  \caption{NDCG@K of DenoiseRank at different denoise-net design on Miscrosoft Web30k, Yahoo! and Istella datasets.}
  \label{fig:denoiseNetLayers}
\end{figure*}

The results (see Figure.\ref{fig:denoiseNetLayers} and Table.\ref{tab:denoiseLayersResult}) show that:
\begin{enumerate}
    \item There is a significant difference between different layers on model performance.
    \item On the web30k dataset, layers=2 performs the best, followed by layers=4, and the performance decreases instead as the layers increase.
    \item On the Yahoo dataset, the model performs significantly better than 6 and 8 when the layers are 2 and 4.
    \item On the istella dataset, the number of layers has no significant effect on model performance.
\end{enumerate}

\begin{table*}[ht]
  \small
  \centering
  \caption{NDCG@K Performance with different denoise network depths on Microsoft Web30K, Yahoo!, and Istella datasets. Best performance per column in bold.}
  \label{tab:denoiseLayersResult}
  \begin{tabular}{@{}lrrrrrrrrr@{}}
    \toprule
    \multirow{2}{*}{Layers} & \multicolumn{3}{c}{Web30K} & \multicolumn{3}{c}{Yahoo!} & \multicolumn{3}{c}{Istella} \\
    \cmidrule(r){2-4} \cmidrule(r){5-7} \cmidrule(r){8-10}
    & @1 & @5 & @10 & @1 & @5 & @10 & @1 & @5 & @10 \\
    \midrule
    2 & \textbf{51.87} & \textbf{52.52} & \textbf{54.60} & 71.37 & 74.06 & 78.42 & 69.00 & 69.10 & 75.69 \\
    4 & 51.45 & 52.08 & 54.03 & \textbf{71.37} & \textbf{74.06} & \textbf{78.42} & \textbf{69.54} & 69.14 & 75.65 \\
    6 & 50.45 & 51.36 & 53.32 & 69.03 & 72.09 & 76.51 & 69.41 & 69.01 & 75.65 \\
    8 & 50.08 & 51.73 & 53.82 & 69.73 & 72.13 & 76.58 & 69.46 & \textbf{69.17} & \textbf{75.80} \\
    \bottomrule
  \end{tabular}
\end{table*}

\subsection{Self Attentions}
\label{sec:SelfAttention}

\begin{figure}[!t]
  \centering
  \includegraphics[width=.5\textwidth]{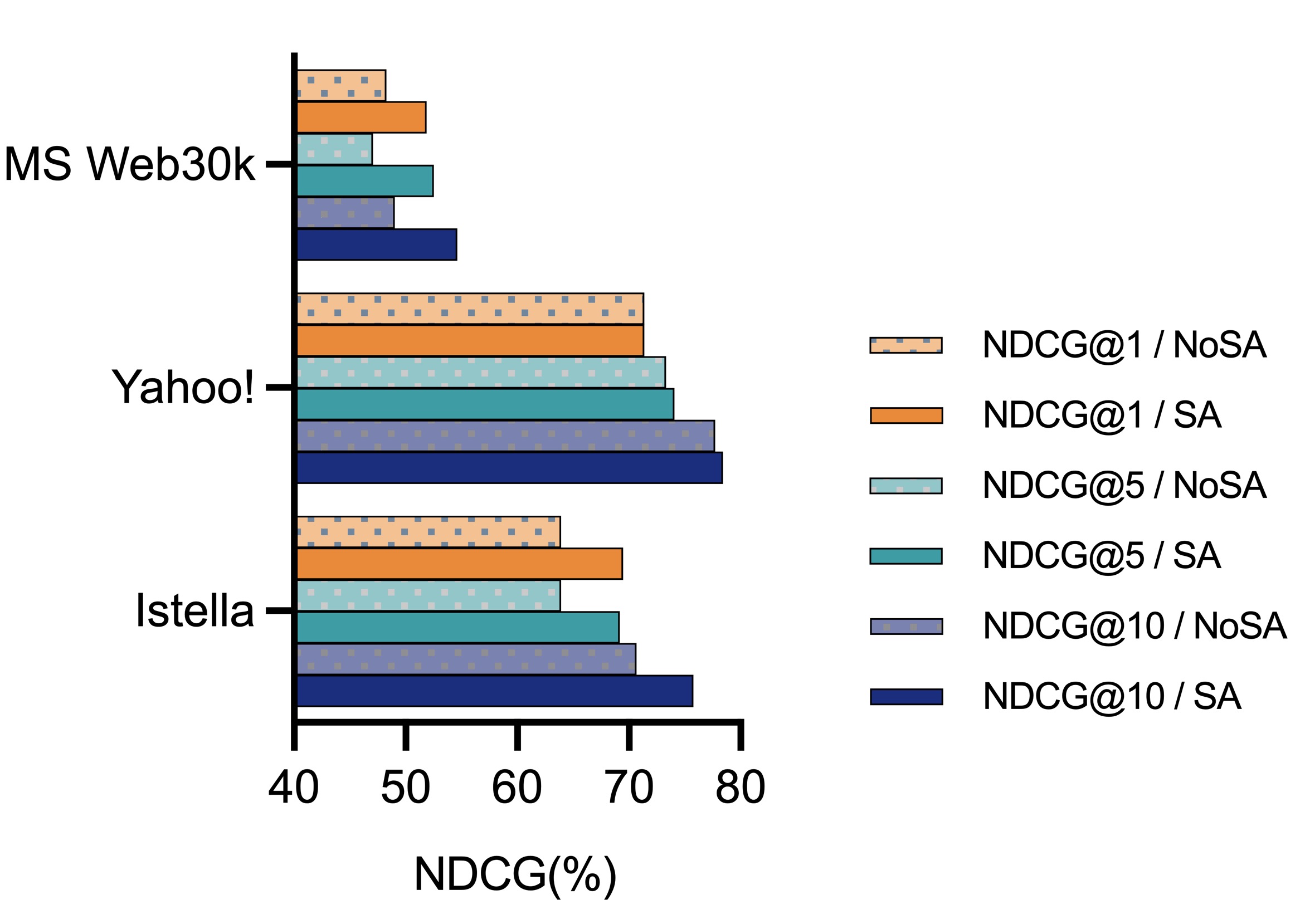}
  \caption{NDCG@K of DenoiseRank without and with self attention on Miscrosoft Web30k, Yahoo! and Istella datasets.}
  \label{fig:selfattentions}
\end{figure}

In recent studies on learning-to-rank \citep{pang2020setrank}\citep{qin2021neural}\citep{buyl2023rankformer}, the self-attention mechanism has been shown to significantly improve ranking results.
To evaluate the effectiveness of self-attention (SA) in DenoiseRank, we conducted experiments with and without Transformer, and the results (see Figure.\ref{fig:selfattentions} and Table.\ref{tab:selfAttentionResult}) show that SA significantly improves the model performance, especially on the MS Web30k and Istella datasets.

\begin{table*}[!t]
  \small
  \centering
  \caption{NDCG@K performance with/without self-attention on Microsoft Web30K, Yahoo!, and Istella datasets. Best performance per column in bold. $\uparrow$ denotes significant improvements.}
  \label{tab:selfAttentionResult}
  \begin{tabular}{@{}lrrrrrrrrr@{}}
    \toprule
    \multirow{2}{*}{Self-Attention} & \multicolumn{3}{c}{Web30K} & \multicolumn{3}{c}{Yahoo!} & \multicolumn{3}{c}{Istella} \\
    \cmidrule(r){2-4} \cmidrule(r){5-7} \cmidrule(r){8-10}
    & @1 & @5 & @10 & @1 & @5 & @10 & @1 & @5 & @10 \\
    \midrule
    Without & 48.25 & 47.08 & 49.00 & 71.37 & 73.35 & 77.70 & 63.91 & 63.91 & 70.65 \\
    With    & \textbf{51.87}\textsuperscript{$\uparrow$} & \textbf{52.52}\textsuperscript{$\uparrow$} & \textbf{54.60}\textsuperscript{$\uparrow$} & \textbf{71.37}\textsuperscript{$\uparrow$} & \textbf{74.06}\textsuperscript{$\uparrow$} & \textbf{78.42}\textsuperscript{$\uparrow$} & \textbf{69.46}\textsuperscript{$\uparrow$} & \textbf{69.17}\textsuperscript{$\uparrow$} & \textbf{75.80}\textsuperscript{$\uparrow$} \\
    \bottomrule
  \end{tabular}
\end{table*}

\section{Diversity}
\label{sec:RankingDiversity}

In inference stage, DenoiseRank randomly samples $Y_T$ from Gaussian noise, causing uncertainty to result of the ranking. Different Gaussian noise labels $Y_T$ may have a different ranking sequence. Compared to DenoiseRank, traditional LTR models may have been trained to rank certainly in various attempts given the same input document features. 

\begin{enumerate}
    \item In real-world information retrieval, the diverse ranked list of items in different search sceneries can be meaningful for at least 3 reasons:
    \item Traditional LTR inclines to rank consistently, which lets the ranking result homogenized and trap users in an information cocoon.
    \item Tapping into the `long-tail ecosystem'. This is a way to boost premium content exposure.
    \item Traditional LTR tends to fall into local optima, DenoiseRank can provide diversity rank results that may be more accurate.
\end{enumerate}

For example:
\begin{enumerate}
    \item In self-media community, diverse ranking can provide premium creative content of the long-tail for users, which can encourage new creators.
    \item Shopping retrieval on the e-Commerce website, we want items with the same relevance scores to have a fair chance to rank higher.
\end{enumerate}

Unfortunately, previous LTR models did not consider uncertainty for ranking and may not rank items diversely.

In this study, we denote diversity in LTR task as: given a query $Q$ and the corresponding documents $D$, run inference by LTR model $M$ times, the number of different ranking sequences is the diversity. For instance, ranking sequence of items $[a,b,c,d,e,f]$ and $[a,b,d,c,f,e]$ are inferenced at different times and causing diversity.

In order to evaluate diversity of our DenoiseRank, we are the first time to introduce a new metric RSD (Ranking Sequence Diversity), formulated as:
\begin{equation}
\small
    RSD@(K,M) = \frac{N}{M}\,,
    \label{eq:RankingSequenceDiversity}
\end{equation}
\noindent where $K$ denotes the top K corresponding documents of the ranking results, $N$ denotes the number of different sequences of items in $M$ times inferred and $N \in [1, M]$. In LTR task, we hope that $RSD@(K,M)$ increases, while $NDCG@K$ does not significantly decrease. There are various metrics that are relevant to ranking diversity, including $Coverage$, $ERR@K$, $\mathrm{Precition-IA}@K$, $DIVERSITY@K$, $\alpha$-NDCG@K etc. However, those metrics are quite different to our $RSD@(K,M)$ in at least the following aspects: (1) $RSD$ considers $M$ times inference and $M$ sequences of items, while traditional diversity metrics focus on single-ranking sequence of items. (2) $RSD$ focus on the diversity of the sequences, while traditional diversity metrics consider the similarity between items in a single sequence. For instance, $DIVERSITY@K$ consider the similarity of pair-wise documents among the top K documents in the result sequence. The higher the metric, the more dissimilar the pair.

We conducted experiments on MS Web30K datasets to investigate the ranking sqeuence deiversity of DenoiseRank and Rankformer which has the similar architecture. We set $K\in\{1,5,10,20\}$ and $M=10$. Then, given a well-trained DenoiseRank model and Rankformer model, we evaluate them on the same test file of Web30K datasets. 
The result of NDCG@K and RSD@(K,M) shown in Table. \ref{tab:RankingSequenceDiversity}, and we find that: 
\begin{enumerate}
    \item Among 10 times inferences, the $RSD$ is 0.11, 0.16, 0.28, 0.64 in the top 1,5,10,20 positions, respectively, showing the ability to rank sequence diversity of DenoiseRank and the Gaussian sampling for $Y_T$ provides uncertainty to rank.
    \item Performance of NDCG@K remains excellent and even slightly increases after repeat inference, which means that our DenoiseRank can produce diverse ranked lists while guarantees reliability of ranking result.
    \item Rankformer did not present the ability to rank in different order, the $RSD$ is 0.1 regardless of the K poisition. It proved our extrapolate that traditional LTR models do not inject uncertainty which results in a static ranking sequence.
    \item Although the NDCG remains excellent on average while enhancing diversity, the deviation results also indicate that isolated extreme cases may occur, leading to either low or high NDCG@K.
\end{enumerate}

According to the above analysis, our DenoiseRank can be applied to areas requiring diverse ranking sequences of items. Our novel metric, RSD, can also be used to evaluate the ranking diversity ability of models in other areas. However, it is important to note that in a few extreme cases, less-related documents may appear at the top. It remains one of the future research questions, with optimising sampling noise being a potential approach 

\begin{figure}
  \centering
  \includegraphics[width=0.98\linewidth]{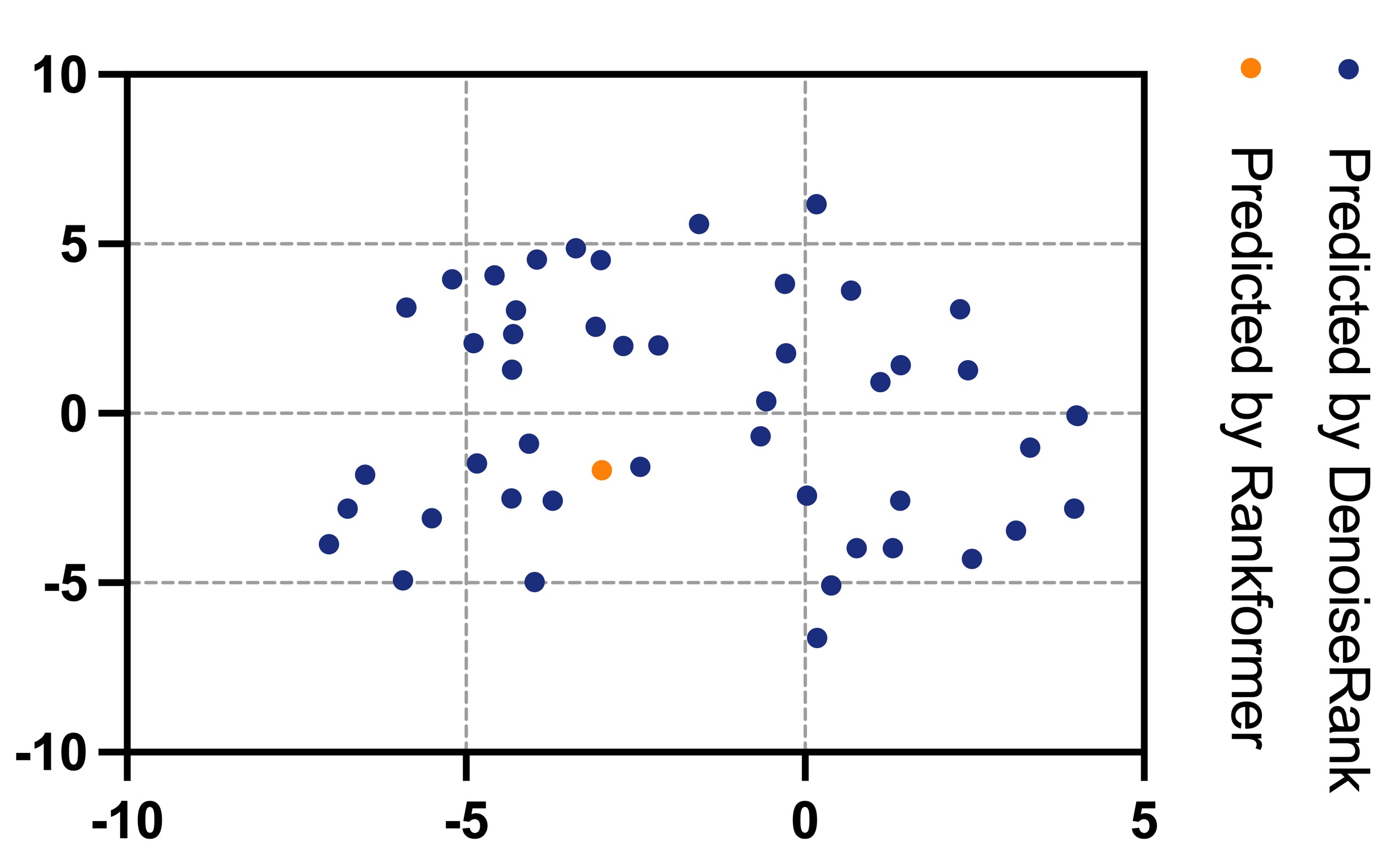}
  \caption{A t-SNE plot shows the diverse ranking sequences on the top 20 predicted in the inference stage of a single query randomly selected. The blue points denote the ranking sequences inferred by DenoiseRank using 100 different $Y_T$ values from Gaussian noise. The orange yellow represents the other sequences predicted by Rankformer in 100 attempts. Testing was conducted on the MS Web30K dataset. }
  \label{fig:diversityRankingtSNE}
\end{figure}

In addition, as our model ranks with diversity, it may not be suitable for those who expect consistent rankings for the same query. Although DenoiseRank can produce ranking results as consistent as possible by sampling the same $Y_t$, in reverse stage, noise is also injected to label and may cause uncertainty.

\begin{table*}[!t]
  \small
  \centering
  \caption{NDCG@K and RSD@(K,M) performance of DenoiseRank and Rankformer on Microsoft Web30K datasets.}
  \label{tab:RankingSequenceDiversity}
  \begin{tabular}{@{}lccccccccc@{}}
    \toprule
    \multirow{2}{*}{Model} & \multirow{2}{*}{M} & \multicolumn{4}{c}{RSD@(K,M)} & \multicolumn{4}{c}{NDCG@K} \\
    \cmidrule(r){3-6} \cmidrule(r){7-10}
    &  & 1 & 5 & 10 & 20 & 1 & 5 & 10 & 20 \\
    \midrule
    \multirow{2}{*}{Rankformer} & 1 & -- & -- & -- & -- & 49.62 & 49.30 & 51.42 & 54.29  \\
                               & 10 & 0.1 & 0.1 & 0.1 & 0.1 & 49.62 & 49.30 & 51.42 & 54.29  \\
    \midrule
    \multirow{2}{*}{DenoiseRank} & 1 & -- & -- & -- & -- & 51.48 & 52.46 & 54.47 & 57.49  \\
                                & 10 & 0.11 & 0.16 & 0.28 & 0.64 & 51.73 & 52.52 & 54.47 & 57.45  \\
    \bottomrule
  \end{tabular}
\end{table*}

\section{Loss Functions}
\label{sec:Lossfunction}

DenoiseRank employs MSE as the loss function in order to predict $Y_0$ at every timestep, see Eq.\ref{eq:denoiserankLossfunctionMSEwithEpsilon}. MSE is also the original loss in DDPMs. Defining suitable ranking losses is an important branch of LTR studies, and there are many versions of loss functions that significantly improve the effectiveness of models. To align with this, we evaluate the performance of DenoiseRank with different losses for ranking. Thus, in this study, we try to find an optimal loss function for DenoiseRank on different datasets.

We consider the following loss functions:
\begin{enumerate}
    \item RMSE: a typical point-wise loss: $L_{\mathrm{RMSE}}(Y, \hat{Y}) = \sqrt{\frac{1}{n}\sum_{i=1}^{n}(Y_i - \hat{Y}_i)^2}$.
    \item RankNet\citep{burges2005learning}: a classic pair-wise loss: $L_{\mathrm{RankNet}}(Y, \hat{Y}) = \sum_{Y_i>Y_j}\log_e(1+e^{\hat{Y}_j-\hat{Y}_i})$.
    \item NDCGLoss$_{2++}$\citep{wang2018lambdaloss}:  a NDCG metric-driven loss functions based on the lambdaLoss 
probabilistic framework: \\
\begin{align}
\small
&L_{\mathrm{NDCGLoss2++}}(Y, \hat{Y})=-\sum\limits_{Y_i>Y_j}\log_2 \nonumber\\
&\sum\limits_{\pi}(\frac{1}{1+e^{-\sigma(\hat{Y}_i-\hat{Y}_j)}})^{(\rho_{ij}+\mu\delta_{ij})|G_i-G_j|}H(\pi|\hat{Y})\nonumber\,,
\end{align}
where $G_i=\frac{2^{y_i}-1}{\mathrm{maxDCG}}$, $\rho_{ij}=|\frac{1}{Di}-\frac{1}{D_j}|$, $\delta_{ij}=|\frac{1}{D_{|i-j|}}-\frac{1}{D_{|i-j|}+1}|$, $D_i=\log_2(1+i)$, and $H(\pi|\hat{Y})$ is a hard assignment distribution of permutations.
    \item ApproxNDCG\citep{qin2010general}\citep{bruch2019revisiting}: a loss that designed to be approximation of NDCG metrics, $L_{\mathrm{ApproxNDCG}}(Y, \hat{Y}) = \frac{1}{Z} \sum_{i=1}^{n} \frac{G(Y_i)}{\log_2(1 + \pi(i))}$, where $Z = -DCG(\pi^*, Y)$, $ G(Y_i)=2^{Y_i}-1$ and $\pi(i)=\frac{1}{2}+\sum_j\mathrm{sigmoid}(\frac{\hat{Y}_j-\hat{Y}_i}{T})$, $T$ is a smooth parameter.
    \item ListNet\citep{cao2007learning}: a classic list-wise loss: $L_{\mathrm{ListNet}}(Y, \hat{Y}) = - \sum^{n}_{i=1}Y_i\log_e\frac{e^{\hat{Y}_i}}{\sum_je^{\hat{Y}_j}}$.
    \item MSE~\citep{ho2020denoising}\citep{nichol2021improved}: a loss function use in DDPMs to predict $x_0$ or $\epsilon$, here we formulate it as $L_{\mathrm{MSE}}(Y, \hat{Y}) = \mathbb{E}[\mid\mid Y - \hat{Y} \mid\mid^2]$
\end{enumerate}

We report the results  based on the best NDCG@10 for different losses. For different loss functions, we use AdamW optimizer and scan learning rate $\in {0.01,0.001,0.0001}$. We try to find the best performance of every loss and report the results based on the NDCG@10. The results are shown in Table.~\ref{tab:lossFunctionResult}, we find that:

\begin{enumerate}
    \item DenoiseRank, when trained with MSE, RMSE and ListNet, achieves first-tier performance and is far superior to the rest.
    \item Though ApproxNDCG improves the performance of neural LTR models in the original papers, it does not seem to work well on DenoisRank, which is implemented from a generative perspective.
    \item DenoiseRank, when trained with ListNet, performs the best on the Web30K dataset. However, for the Yahoo! and Istella datasets, training with MSE loss is the best choice.
\end{enumerate}

\begin{table*}[!t]
  \small
  \centering
  \caption{NDCG@K performance of DenoiseRank with different loss functions on Microsoft Web30K, Yahoo!, and Istella datasets. Best performance per column in bold.}
  \label{tab:lossFunctionResult}
  \begin{tabular}{@{}lrrrrrrrrr@{}}
    \toprule
    \multirow{2}{*}{Loss} & \multicolumn{3}{c}{Web30K} & \multicolumn{3}{c}{Yahoo!} & \multicolumn{3}{c}{Istella} \\
    \cmidrule(r){2-4} \cmidrule(r){5-7} \cmidrule(r){8-10}
    & @1 & @5 & @10 & @1 & @5 & @10 & @1 & @5 & @10 \\
    \midrule
    RMSE              & 50.48 & 51.41 & 53.43 & 70.64 & 73.30 & 77.84 & 69.26 & \textbf{69.32} & 75.80 \\
    RankNet           & 43.66 & 45.84 & 48.56 & 56.65 & 66.96 & 73.18 & 51.11 & 57.25 & 65.83 \\
    NDCGLoss\textsubscript{2++} & 43.01 & 47.68 & 50.57 & 66.48 & 72.18 & 76.98 & 56.00 & 59.96 & 67.55 \\
    ApproxNDCG        & 24.46 & 29.21 & 33.72 & 60.24 & 64.67 & 70.64 & 33.01 & 42.21 & 52.48 \\
    ListNet           & \textbf{51.87} & \textbf{52.52} & \textbf{54.60} & 70.81 & 73.82 & 78.35 & 68.75 & 69.05 & 75.58 \\
    MSE               & 51.20 & 51.73 & 53.77 & \textbf{71.37} & \textbf{74.06} & \textbf{78.42} & \textbf{69.4}6 & 69.17 & \textbf{75.80} \\
    \bottomrule
  \end{tabular}
\end{table*}

\section{Other Metrics}
\label{sec:Othermetric}

In order to evaluate our denoiseRank  fully, we use another 4 types of ranking metrics, including Expected Reciprocal Rank (ERR), Mean Average Precision (MAP), Mean Reciprocal Rank(MRR) and Precision. We reported results at ranks 1,3,5,10,20 and the total rank (denoted as ``ALL''). The results are shown in Table.~\ref{tab:Othermetric}, and the results further confirm the effectiveness of our models.

\begin{table*}[!t]
  \small
  \centering
  \caption{ERR, MRR, MAP and precision of DenoiseRank on Microsoft Web30K, Yahoo!, and Istella datasets.}
  \label{tab:Othermetric}
  \begin{tabular}{@{}llrrrllrrr@{}}
    \toprule
    \cmidrule(lr){1-5} \cmidrule(l){6-10}
    \multirow{2}{*}{Metric} & \multirow{2}{*}{K} & \multicolumn{3}{c}{Dataset} & \multirow{2}{*}{Metric} & \multirow{2}{*}{K} & \multicolumn{3}{c}{Dataset} \\
    \cmidrule(r){3-5} \cmidrule(l){8-10}
    & & Web30K & Yahoo! & Istella & & & Web30K & Yahoo! & Istella \\
    \midrule
    \multirow{5}{*}{ERR} 
    & 1  & 26.53 & 34.41 & 61.53 & \multirow{5}{*}{MAP}  & 1  & 78.07 & 87.13 & 94.64 \\
    & 5  & 36.77 & 43.90 & 73.79 &                    & 5  & 81.55 & 89.16 & 95.19 \\
    & 10 & 38.55 & 45.34 & 74.34 &                    & 10 & 78.76 & 87.91 & 93.14 \\
    & 20 & 39.28 & 45.72 & 74.40 &                    & 20 & 74.89 & 86.72 & 90.27 \\
    & ALL& 39.57 & 45.78 & 74.40 &                    & ALL& 63.91 & 85.75 & 88.36 \\
    \midrule
    \multirow{5}{*}{MRR} 
    & 1  & 78.07 & 87.13 & 94.64 & \multirow{5}{*}{Precision} & 1  & 78.07 & 87.13 & 94.64 \\
    & 5  & 84.36 & 90.56 & 96.77 &                    & 5  & 72.86 & 83.59 & 89.38 \\
    & 10 & 84.64 & 90.69 & 96.79 &                    & 10 & 69.25 & 81.25 & 80.44 \\
    & 20 & 84.73 & 90.70 & 96.79 &                    & 20 & 64.17 & 78.81 & 55.41 \\
    & ALL& 84.75 & 90.71 & 96.79 &                    & ALL& 44.97 & 75.30 & 12.81 \\
    \bottomrule
  \end{tabular}
\end{table*}

\section{Effeciency}
\label{sec:effeciency}

DenoiseRank is a diffusion-based model that need to inference by mutiple reverse steps, which may introduced computational overhead, it is acceptable if pretty better results can be obtained with a small number of reverse steps. We therefore compared the inference times of the baselines experimentally, and then testing DenoiseRank's inference time and metrics at different reverse steps.All the experiments are conducted on an NVIDIA GeForce RTX 3090 GPU and Intel Xeon CPU Gold 6226R. The results are as follows:

\begin{enumerate}
    \item As shown in Table \ref{tab:Infer-time-compare}, the inference time of the DenoiseRank with a smaller reverse step (4 steps) is close to that of NeuralNDCG, but less than the Rankformer. Their inference times are not vastly different, due to the fact that they have similar model architectures.\\
    \item According to Table \ref{tab:Infer-time-diff-step}, DenoiseRank's inference time increases quickly as the reverse step increases; however, NDCG performance fluctuates(Table \ref{tab:NDCG-diff-step}). Therefore, it is necessary to strike a balance between time and performance and to use the appropriate time step in practice.\\
    \item Compared to neural network-based models, tree-based models require very little inference time due to their computational complexity and algorithmic properties. Therefore, tree-based models should be used in systems requiring a high response speed, and DenoiseRank should be used in scenarios involving complex representations and user feedback distribution.
    \item In Table \ref{tab:Infer-time-compare}, we report the training time of the baselines. The time processed by the tree-based models is significantly lower than that by the neural network-based models, which is due to their more light-weight model structure and size. Additionally, DenoiseRank consumes medium training time in the neural network model. Although the forward process requires multiple steps, we can perform the noise addition in one step without consuming much extra time by using tricks such as reparameterization.
\end{enumerate}

Therefore, inference and training time will not be a bottleneck for optimisation in DenoiseRank. Despite the fact that tree-based models require significantly less time, DenoiseRank did not demonstrate vastly different efficiency to other neural-based models and can be applied to scenarios involving complex data.

\begin{table*}[!t]
\centering
\small  
\caption{Total inference time(s) at different reverse steps of DenoiseRank on Yahoo datasets.}  
\label{tab:Infer-time-diff-step}  
\begin{tabular}{@{}lccccccccc@{}}  
\toprule
\multicolumn{1}{l}{\textbf{Step}} & \textbf{2} & \textbf{4} & \textbf{8} & \textbf{16} & \textbf{32} & \textbf{64} & \textbf{128} & \textbf{256} & \textbf{512} \\
\midrule
Times(s)       & 4.75 & 5.28 & 6.42 & 8.76 & 13.44 & 23.57 & 44.11 & 85.04 & 166.73 \\
\bottomrule
\end{tabular}
\end{table*}

\begin{table*}[!t]
\centering
\small  
\caption{NDCG@10 of DenoiseRank at different reverse steps.}  
\label{tab:NDCG-diff-step}  
\begin{tabular}{@{}lccccccccc@{}}  
\toprule
\multicolumn{1}{l}{\textbf{Step}} & \textbf{2} & \textbf{4} & \textbf{8} & \textbf{16} & \textbf{32} & \textbf{64} & \textbf{128} & \textbf{256} & \textbf{512} \\
\midrule
MS Web30K       & 54.59 & 54.60 & 54.62 & 54.61 & 54.57 & 54.61 & 54.58 & 54.61 & 54.59 \\
Yahoo!       & 78.41 & 78.40 & 78.42 & 78.38 & 78.37 & 78.36 & 78.38 & 78.39 & 78.38 \\
Istella       & 75.80 & 75.81 & 75.82 & 75.79 & 75.80 & 75.79 & 75.80 & 75.84 & 75.82
 \\
\bottomrule
\end{tabular}
\end{table*}

\begin{table*}[!t]
\centering
\small  
\caption{Inference and training times of baselines on Yahoo datasets.}  
\label{tab:Infer-time-compare}  
\begin{tabular}{@{}lccccc@{}}  
\toprule
\multicolumn{1}{l}{\textbf{Stage}} & \textbf{DenoiseRank} & \textbf{Rankformer} & \textbf{NeuralNDCG} & \textbf{$\lambda\mathrm{MART}_{\mathrm{RankLib}}$} & \textbf{$\lambda\mathrm{MART}_{\mathrm{GBM}}$} \\
\midrule
Inference(ms)       & 0.92 & 3.35 & 0.68 & 1.34 & 0.04  \\
Training(s)      & 13.88  & 15.46 & 11.23 & 1.38  & 0.20   \\  
\bottomrule
\end{tabular}

\medskip
\footnotesize
\textit{Note}: 'Inference (ms)' denotes the average time taken to inference each query. \\'Training (s)' denotes the average time cost of each training batch.
\\ DenoiseRank inference at 4 reverse steps.
\end{table*}

\end{document}